\documentclass[aps,nofootinbib,prd,superscriptaddress,tightenlines,notitlepage,twocolumn,showpacs,floatfix]{revtex4-1}
\usepackage[colorlinks]{hyperref}
\usepackage{tabularx}
\usepackage{bm}
\usepackage{graphicx}
\usepackage{amssymb,bm,tensor}
\usepackage{xcolor}
\usepackage{amsmath}
\usepackage[varg]{txfonts}
\usepackage{enumerate}
\usepackage[normalem]{ulem}
\usepackage{bm}
\usepackage[OT1]{fontenc}
\usepackage{threeparttable}
\usepackage{cleveref}

\usepackage{scalerel,stackengine}
\stackMath
\newcommand\reallywidehat[1]{%
\savestack{\tmpbox}{\stretchto{%
  \scaleto{%
    \scalerel*[\widthof{\ensuremath{#1}}]{\kern-.6pt\bigwedge\kern-.6pt}%
    {\rule[-\textheight/2]{1ex}{\textheight}}
  }{\textheight}%
}{0.5ex}}%
\stackon[1pt]{#1}{\tmpbox}%
}

\begin{document}

\title{Linear analysis of I-C-Love universal relations for neutron stars}

\date{\today}

\author{Zexin Hu}
\affiliation{Department of Astronomy, School of Physics, Peking University, Beijing 100871, China}
\affiliation{Kavli Institute for Astronomy and Astrophysics, Peking University, Beijing 100871, China}
\affiliation{Theoretical Astrophysics, Eberhard Karls University of T\"ubingen, 72076 T\"ubingen, Germany}

\author{Yong Gao}
\affiliation{Max Planck Institute for Gravitational Physics (Albert Einstein Institute), 14476 Potsdam, Germany}

\author{Lijing Shao}
	\email{lshao@pku.edu.cn}
\affiliation{Kavli Institute for Astronomy and Astrophysics, Peking University, Beijing 100871, China}
\affiliation{National Astronomical Observatories, Chinese Academy of Sciences, Beijing 100012, China}

\begin{abstract}

Neutron stars (NSs) are excellent laboratories for testing gravity theories as
they are strongly self-gravitating bodies and have rich observational phenomena.
However, strong-field gravity effects in NS could be degenerate with their equation of
state (EOS) which is largely unknown. Fortunately, there exist the so-called
universal relations among the NS macroscopic quantities that are found to be
insensitive to the underlying EOS.  Studying the origin of these relations can
lead to a better understanding of NSs and the gravitational interaction.  We
develop a new perspective of view to analyze the I-C and I-Love universal
relations for NSs. At the linear order, we separate the deviation 
of the universal relations into two factors that are multiplied together. One is the 
EOS difference while the other factor only depends on the background star structure.
The smallness of the second factor that does not depend on the EOS difference then 
indicates the origin of the universality. We discuss the validity of our linear approximation
when considering the difference among realistic EOSs. Our study can be regarded
as a new frame for quantitative representation of the universality and may
provide new insights to the universal relations of NSs.

\end{abstract}

\maketitle


\allowdisplaybreaks

\section{Introduction}

Neutron stars (NSs) are extremely compact objects in the Nature. Thanks to their
strong self-gravity, large density in the inner region, and abundant
observational phenomena, they are ideal laboratories for studying fundamental
physics, including the strong-field gravity and nuclear
physics~\cite{Lattimer:2000nx, Shao:2022izp, Freire:2024adf}. For example, the
mass-radius relationship of NSs strongly depends on the supranuclear matter
equation of state (EOS), which is the relation between the matter density $\rho$
and pressure $p$. Thus, independently measuring the mass  $M$ and radius $R$ of
a NS  can provide valuable constraints on the EOS at the high density regime
that is unachievable with terrestrial nuclear experiments. The observations of
heavy pulsars with masses $\gtrsim 2\,M_\odot$ have already ruled out those EOSs
that are too ``soft'' for supporting a large mass NS in the general relativity
(GR)~\cite{Fonseca:2021wxt, Demorest:2010bx, Antoniadis:2013pzd}. Current X-ray
observations with the NICER satellite can measure the mass and radius of a
pulsar to $\sim 10\%$ accuracy and have also been used to constrain the
EOS~\cite{Riley:2019yda, Raaijmakers:2019qny}. 

However, the mass-radius relationship of NSs can be affected by the underlying
gravity theory as well.  A well-known example is the scalar-tensor theory, while
passing all weak-field gravity tests in some parameter
space~\cite{Will:2018bme}, predicts large deviation in the strong-field regime
with a phenomenon called spontaneous scalarization~\cite{Damour:1993hw,
Damour:1996ke, Doneva:2022ewd}. Binary NSs were used as excellent testbeds, notably
with the observational absence of the gravitational dipole radiation that is
otherwise predicted in the scalar-tensor gravity theory~\cite{Freire:2012mg,
Shao:2017gwu, Zhao:2022vig}. But still, considering a massive scalar field can
evade these dipole-radiation constraints~\cite{Ramazanoglu:2016kul, Doneva:2016xmf, 
Hu:2020ubl, Xu:2020vbs, Hu:2021tyw}.

The degeneracy between the strong-field gravity effects and the nuclear matter
EOS in NSs causes a challenge for studying both aspects~\cite{Shao:2019gjj}. A
test of strong-field gravity can only be done if we assume that we know the
underlying EOS or systematically taken into account its large uncertainties, and
vise versa. One way to break this degeneracy is provided by the finding of the
so-called universal relations among some NS properties~\cite{Yagi:2013bca,
Yagi:2013awa, Pappas:2013naa, Chakrabarti:2013tca, Gao:2023mwu, Luk:2018xmt,
Shao:2022koz}.  These relations, for example, the well-known I-Love-Q relation
for the dimensionless NS moment of inertia $\tilde{I}$, tidal Love number $\tilde{\Lambda}$ and spin induced
quadrupole moment $\tilde{Q}$, are largely insensitive to the EOSs but sometimes do depend
on the gravity theories or even not universal in theories beyond
GR~\cite{Yagi:2016bkt}. Thus, independent measurement of these quantities will
enable us to have EOS independent test of gravity theories. In this regard, the
I-Love relation will be especially interesting as the tidal Love number of NS
can be measured with gravitational wave 
observations~\cite{LIGOScientific:2018cki}, while it is expected to measure the
NS moment of inertia with the timing observation of the Double Pulsar
J0737$-$3039A/B~\cite{Hu:2020ubl, Kramer:2021jcw}.

Besides the interesting applications of these universal relations, studying
themselves also provides us better understanding of NSs and gravity theories.
The origin of the universality, which can be as good as $\lesssim 1\%$ level in
the I-Love-Q relations, is still in discussion. In literature, it is proposed at
the beginning that the universality may come from the similarity of the
different EOSs in the low density regime or it is the remnants of the no-hair
theorem~\cite{Yagi:2013awa,Yagi:2013bca}.  Considering that some relations like
I-Love-Q even hold among NSs and quark stars (QSs), which are composed with
quark matter that has a completely different EOS behavior in the low density
regime, it is then proposed that the universality may come from the emergent 
self-similarity of isodensity contours inside NSs~\cite{Yagi:2014qua} or the
realistic EOS is ``close'' to the incompressible limit~\cite{Sham:2014kea,
Katagiri:2025qze}.  There are also analytical calculations of these relations
based on series expansion of the star's compactness and assuming an
incompressible EOS~\cite{Chan:2014tva} or Tolman VII model that
phenomenologically describes the density profile inside NSs~\cite{Jiang:2020uvb,
Lowrey:2024anh}.

Most of the studies on analyzing the NS universal relations take some specific 
parametrization of the EOS or assumptions of the NS configurations, and the 
relations are also often expanded around the Newtonian limit. These attempts may
reveal the possible key ingredient that preserves the universality, but also may
lead to questions that why these configurations hold for various EOSs. Though
there are a limited number of studies using different approaches,
\citet{Chan:2015iou} showed that the I-Love relation in the incompressible limit
is stationary at the leading order of EOS variation and star's compactness.

In this work, we analyze the NSs' I-C and I-Love universal relations from a new
perspective of view. Rather than parametrizing or expanding the EOSs or NS
structures, we consider the linear response of the universal relation under {\it
any} EOS perturbations. With this approach, we are able to separate the
deviation of the universal relation caused by EOS perturbations into two parts 
that are multiplied together:
one is the difference in the EOSs and the other factor only depends on the 
background star solution.  As one can have almost arbitrary EOS 
difference if the universal relation holds, the smallness of the second factor fully
accounts for the universality. Our approach also does not assume an expansion in
the star's compactness thus is applicable for both Newtonian and GR
configurations.

The remaining part of this paper is organized as follows. In Sec.~\ref{sec:I-C}
we  display the basic equations for calculating the I-C universal relations, and
we introduce our concepts and approaches in Sec.~\ref{sec:local U}. We apply our
approach to a simple case that can be solved analytically in
Sec.~\ref{sec:example}, and we apply the approach for realistic EOSs in
Sec.~\ref{sec:real}. Section~\ref{sec:I-Love} shows the results for I-Love
relation. We also discuss the validity of our linear approximation when
considering the difference between realistic EOSs in Sec.~\ref{sec:linear}.
Finally, we summarize our work in Sec.~\ref{sec:sum}. Throughout this work, we
use the geometric units of $G=c=1$.

\section{I-C universal relation}\label{sec:I-C}

Here we briefly review the I-C universal relation and the basic equations for
the relevant NS properties. Though it is known that the I-C relation does not
hold as good as the I-Love relation, we first focus on it for its simplicity. We
further extend our analysis to I-Love relation in later sections and compare the
difference of them.

The ordinary differential equations that govern a relativistic, non-rotating
star made of perfect fluid are the well-known Tolman-Oppenheimer-Volkov (TOV) 
equations~\cite{Oppenheimer:1939ne,Tolman:1939jz},
\begin{eqnarray}
  \frac{{\rm d}r}{{\rm d}p}&=&-\frac{1}{p+\rho}\frac{r^2(1-2m/r)}{m+
  4\pi r^3 p}\,,\label{eq:drdp}\\
  \frac{{\rm d}m}{{\rm d}p}&=&-\frac{\rho}{p+\rho}\frac{4\pi r^4(1-2m/r)}
  {m+4\pi r^3 p}\,,
\end{eqnarray}
where $r$ is the circumferential radius and 
\begin{equation}
  m(r)=\int_0^r 4\pi r'^2 \rho(r') {\rm d}r'\,,
\end{equation}
can be regarded as the gravitational mass enclosed within radius $r$.  Energy
density $\rho$ and pressure $p$ are related by the so-called EOS,
\begin{equation}
  \rho=\rho(p)\,.
\end{equation}
In the above equation, we have assumed that the density is only a function of
the pressure, or to say, we have adopted a cold EOS. The temperature effects are
more important for newly born NSs and do affect the universal relations relating
the post-merger gravitational wave spectrum with the NS
structures~\cite{Raithel:2022orm,Han:2025pho}. Nevertheless, for old NSs with
stable structures, they are usually cold relative to their Fermi temperature and
the above temperature-independent EOS is sufficient.

We calculate the moment of inertia of a NS with the slow rotation 
approximation~\cite{Hartle:1967he, Hartle:1968si, Gao:2021uus}. It can be shown
that, in this approximation, the moment of inertia of the star also can be
described by a single ordinary differential equation~\cite{Grigotian:1997}\,,
\begin{equation}
  \frac{{\rm d}i}{{\rm d}p}=-\frac{2}{3}\frac{4\pi r^6}{m+4\pi r^3 p}
  \left(1-\frac{5}{2}\frac{i}{r^3}+\frac{i^2}{r^6}\right)\,,
\end{equation}
where $i$ can be regarded as the moment of inertial contributed by the matter
inside radius $r(p)$. However, due to the relativistic effect, the above
equation is nonlinear for $i$ and there is no superposition principle as in
Newtonian gravity. There are studies of the universal relations for fast or
differential rotating NSs~\cite{Doneva:2013rha, Pappas:2013naa,Bretz:2015rna}.
However, the calculation of moment of inertia in these situations can be much
more complex and we leave it for future studies.

With the above equations, the radius, mass and moment of inertia of the star are
defined to be 
\begin{equation}
  R=r |_{p=0} \,,\quad M=m |_{p=0} \,,\quad I=i|_{p=0} \,,
\end{equation}
where $p$ vanishes at the boundary of the star.

\begin{figure}[t]
  \centering
  \includegraphics[width=0.45\textwidth]{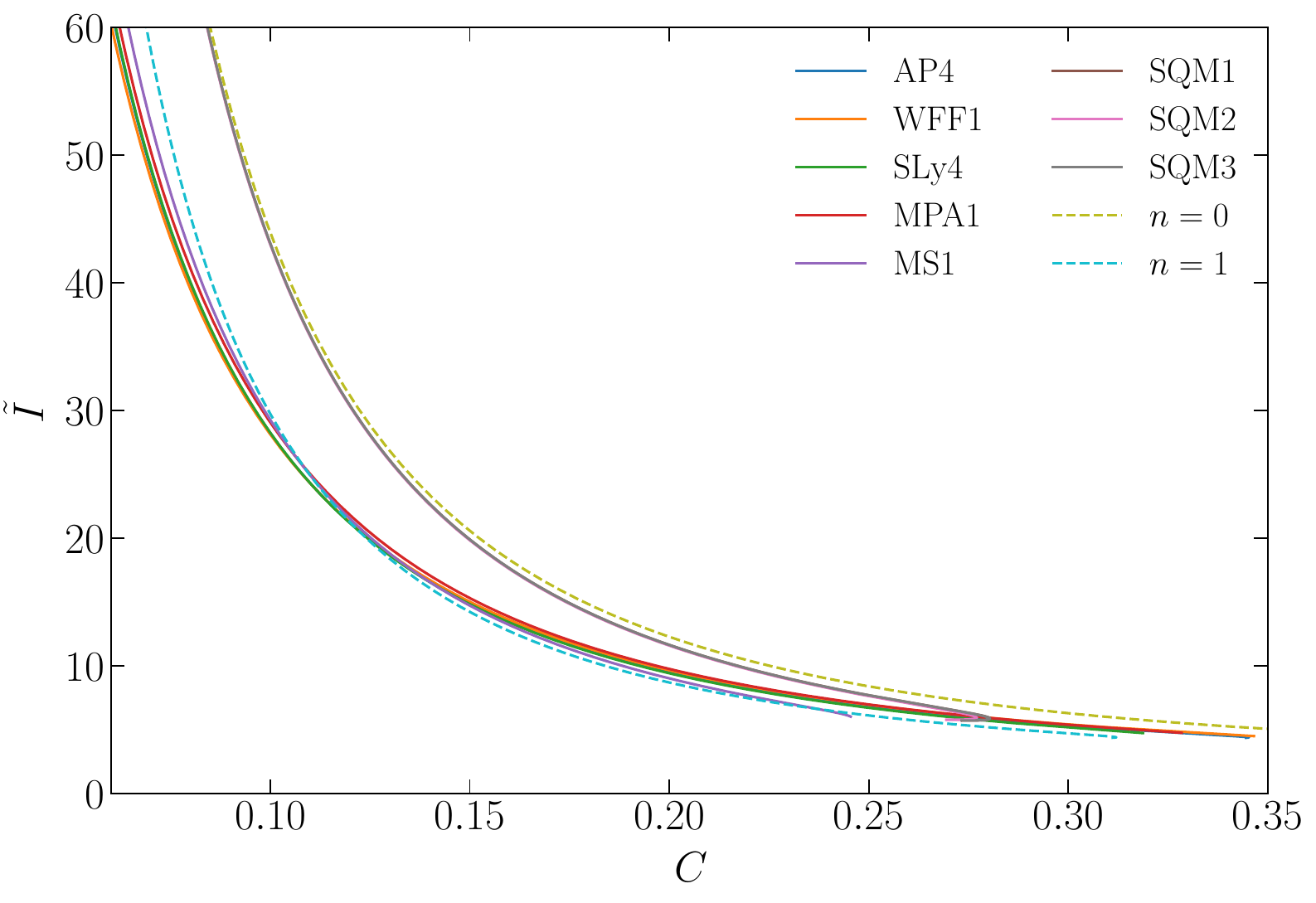}
  \caption{\label{fig:universal_I_C} The I-C relation for NSs and QSs with
  realistic EOSs as well as polytropic EOSs.}
\end{figure}

In Fig.~\ref{fig:universal_I_C} we show the I-C universal relation for several
realistic EOSs including NSs and QSs as well as polytropic EOSs with 
$\rho(p)=\rho_n(p/\rho_n)^{n/(n+1)}$. The dimensionless moment of inertial is
defined to be $\tilde{I}\equiv I/M^3$ and the compactness is
$C=M/R$~\cite{Yagi:2013bca, Yagi:2013awa}.  As we focus on the dimensionless
relation, the dimensionful quantity $\rho_n$ dose not affect the final relation
and we choose it to be $1$ as the unit. From the figure one can see that for
realistic EOSs, the relation holds well to the level of several percentages.
However, this relation has different branches for NSs and QSs. As expected, the
polytropic EOS with $n=1$ is a good approximation for realistic NS EOSs while
$n=0$ can describe the EOSs of QSs.

\section{Linear response of the I-C relation}\label{sec:local U}

In this section, we describe the approach we used to calculate the linear
response of arbitrary EOS perturbations. Starting from any given EOS, say,
$\rho(p)$, and a given central pressure, $p_0$, we may solve the structure
equations given in the previous section with the initial condition that 
\begin{equation}
  p=p_0\,,\quad r=0\,,\quad m=0\,,\quad i=0\,,
\end{equation}
to get a background solution denoted by 
\begin{equation}
  r(p)\,,\quad m(p)\,,\quad i(p)\,,
\end{equation}
and thus $R$, $M$, $I$ for the star at $p=0$. With the background solution, 
considering another EOS given by
\begin{equation}
  \rho'(p)=\rho(p)+\delta\rho(p)\,,
\end{equation} 
but with the same initial condition, we can write the corresponding NS structure
to be
\begin{eqnarray}
  r'(p)&=&r(p)+\delta r(p)\,,\\
  m'(p)&=&m(p)+\delta m(p)\,,\\
  i'(p)&=&i(p)+\delta i(p)\,.
\end{eqnarray}

Assuming $\delta\rho(p)$ is an EOS perturbation satisfying $\delta\rho(p)/\rho(p)\ll 1$ for $p\in[0,p_0]$, 
one can then obtain the linearized differential equations for $\delta r$, $\delta m$ and
$\delta i$ by substituting the above equations to the NS structure equations and 
only keeping the leading order terms. The linearization also requires the variable set $\delta r(p)/r(p)$, 
$\delta m(p)/m(p)$ and $\delta i(p)/i(p)$ to be small, which is in general guaranteed by the assumption 
$\delta\rho(p)/\rho(p)\ll 1$. Note that at $p=p_0$ the condition $r=m=i=0$ is only a coordinate singularity. At the center of the 
star, one can solve the TOV equations to obtain
\begin{equation}
  r(p)=\sqrt{\frac{3}{2\pi}\frac{p_0-p}{(\rho_0+p_0)(\rho_0+3p_0)}}+O(p_0-p)^{3/2}\,,
\end{equation}
where $\rho_0=\rho(p_0)$. Therefore, 
\begin{equation}
  \left.\frac{\delta r}{r}\right|_{p=p_0}=-\frac{\rho_0+2p_0}{(\rho_0+p_0)(\rho_0+3p_0)}\delta\rho(p_0)\,,
\end{equation}
which is also small if $\delta\rho(p_0)/\rho(p_0)\ll 1$; it is similar for $\delta m/m$ and $\delta i/i$.

The linearized equations for the $\delta$-quantities read
\begin{equation}\label{eq:drho}
  \frac{{\rm d}}{{\rm d}p}
  \begin{bmatrix}
    \delta r\\
    \delta m\\
    \delta i
  \end{bmatrix}
  =\bm{A}(p)
  \begin{bmatrix}
    \delta r\\
    \delta m\\
    \delta i
  \end{bmatrix}
  +\bm{S}(p)\delta\rho(p)\,,
\end{equation}
where $\bm{A}(p)$ and $\bm{S}(p)$ are matrix and column vector depending on the
background solution. We present the formulae of $\bm{A}(p)$ and $\bm{S}(p)$ here
for later numerical calculations,
\begin{widetext}
\begin{align}\label{eq:Ap}
  \bm{A}(p) &=
  \begingroup\renewcommand{\arraystretch}{2.6}
  \begin{bmatrix}
    \dfrac{2}{p+\rho}\dfrac{m^2+2\pi r^4p-mr-8\pi r^3 p m}{(m+4\pi r^3p)^2}& \quad
    \dfrac{1}{p+\rho}\dfrac{r^2(1+8\pi r^2 p)}{(m+4\pi r^3p)^2}& \quad 0\\
    \dfrac{\rho}{p+\rho}\dfrac{8\pi r^2(3m^2-2mr-2\pi r^4 p)}{(m+4\pi r^3p)^2}& \quad
    \dfrac{\rho}{p+\rho}\dfrac{4\pi r^4(1+8\pi r^2 p)}{(m+4\pi r^3p)^2}& \quad 0\\
    -\dfrac{4\pi r^2(4mr^3+8\pi r^6 p-5im-8\pi p i^2)}{(m+4\pi r^3p)^2}& \quad
    \dfrac{8\pi r^6(1-5i/2r^3+i^2/r^6)}{3(m+4\pi r^3p)^2}& \quad
    \dfrac{4\pi(5r^3-4i)}{3(m+4\pi r^3p)}
  \end{bmatrix}\,, 
    \endgroup \\
  \bm{S}(p) &=
  \begingroup\renewcommand{\arraystretch}{2.4}
  \begin{bmatrix}
    \dfrac{1}{(p+\rho)^2}\dfrac{r^2(1-2m/r)}{m+4\pi r^3 p}\\
    -\dfrac{p}{(p+\rho)^2}\dfrac{4\pi r^4 (1-2m/r)}{m+4\pi r^3 p}\\
    0
  \end{bmatrix}\,.
  \endgroup
\end{align}
\end{widetext}
The initial condition for the perturbation equation, Eq.~(\ref{eq:drho}), reads,
\begin{equation}
  p=p_0\,,\quad \delta r=0\,,\quad \delta m=0\,,\quad \delta i=0\,.
\end{equation}

A further observation leads to some intuitive simplification of the problem.
Note that the perturbation equation~(\ref{eq:drho}) is linear, thus a natural
and complete choice of $\delta \rho(p)$ is a delta function that $\delta
\rho(p)=\delta(p-p_1)$ with $0<p_1\leq p_0$. Considering this kind of EOS
perturbation, Eq.~(\ref{eq:drho}) is simplified into
\begin{equation}\label{eq:pert}
  \frac{{\rm d}}{{\rm d}p}
  \begin{bmatrix}
    \delta r\\
    \delta m\\
    \delta i
  \end{bmatrix}
  =\bm{A}(p)
  \begin{bmatrix}
    \delta r\\
    \delta m\\
    \delta i
  \end{bmatrix}\,,
\end{equation}
but with initial condition 
\begin{equation}
  p=p_1\,,\quad 
  \delta r=-S_1(p_1)\,,\quad 
  \delta m=-S_2(p_1)\,,\quad 
  \delta i=0\,,
\end{equation}
where we have defined $\bm{S}(p)= \big[S_1(p),S_2(p),0 \big]^\intercal$.

Before further proceeding, here we note that the dimension of the delta function source 
$\delta\rho=\delta(p-p_1)$ is $p^{-1}$ instead of $\rho$ according to the definition of delta function. 
Therefore, the corresponding solution like $\delta r$ has a dimension of $r/p\rho$ instead of $r$, which 
might be less intuitive. 
Nevertheless, physical quantities that are obtained by applying superposition principle have the 
correct dimension as shown below.

Solving the above equation analytically or numerically, one can obtain the 
perturbation quantities as a function of $p$. We denote the solution as 
\begin{equation}
  \delta r(p;p_1)\,,\quad 
  \delta m(p;p_1)\,,\quad 
  \delta i(p;p_1)\,,
\end{equation}
where $p_1$ reminds that the solution is for $\delta \rho(p)=\delta(p-p_1)$.

What we are interested in is the change in the star's radius, mass and moment of
inertial caused by an arbitrary EOS perturbation. As we have linearized the 
equations, we can apply the superposition principle to calculate; for example, 
\begin{equation}\label{eq:inte_R}
  \delta R=\int_0^{p_0}\delta R(p_1) \delta\rho(p_1){\rm d}p_1\,,
\end{equation}
where $\delta R(p_1)\equiv \delta r(p=0;p_1)$. The deviation of the universal 
relation then can be characterized by the following combination,
\begin{equation}\label{eq:DII_C}
  \frac{{\rm D}\tilde{I}}{\tilde{I}}\equiv\frac{\delta\tilde{I}-K\delta C}
  {\tilde{I}}\,,
\end{equation}
where
\begin{equation}
	\delta \tilde{I}/\tilde{I}=\delta I/I-3\delta M/M^3 \,, \quad
	\delta C/C= \delta M/M-\delta R/R \,,
\end{equation}
and $K\equiv {\rm d}\tilde{I}/{\rm d}C$ is the ``true'' slope of the universal
solution.  Notice that it is similar to Eq.~(6.30) in Ref.~\cite{Chan:2015iou}
but we do not expand to the leading order of $C$.  However, due to the fact that
the relation is only approximately universal, $K$ can be calculated with the
background EOS.

Equation~(\ref{eq:DII_C}) is a linear combination of the perturbation 
quantities, thus it also adopts a superposition form 
\begin{equation}\label{eq:pert_integral}
  \frac{{\rm D}\tilde{I}}{\tilde{I}}=\int_0^{1}p_0\rho_0
  \frac{{\rm D}\tilde{I}(p_1)}{\tilde{I}}\frac{\delta \rho(p_1)}{\rho_0}
  {\rm d}x\,,
\end{equation}
where $x=p_1/p_0$, and ${\rm D}\tilde{I}(p_1)/\tilde{I}$ is the corresponding 
quantity for a delta function like EOS perturbation defined before. 
$\rho_0$ can be any number that makes these combination dimensionless, and we 
choose it to be $\rho_0=\rho(p_0)$ as it seems to be the most natural. The
dimensionless factor $p_0\rho_0{\rm D}\tilde{I}(p_1)/\tilde{I}$ then fully
characterizes the linear response of the I-C relation to an arbitrary EOS
perturbation at a given compactness. As a function of $p_1$, it will also tell
us which part of the NS is most sensitive to the EOS perturbation and thus
contributes the most.

\begin{figure}[t]
  \centering
  \includegraphics[width=0.45\textwidth]{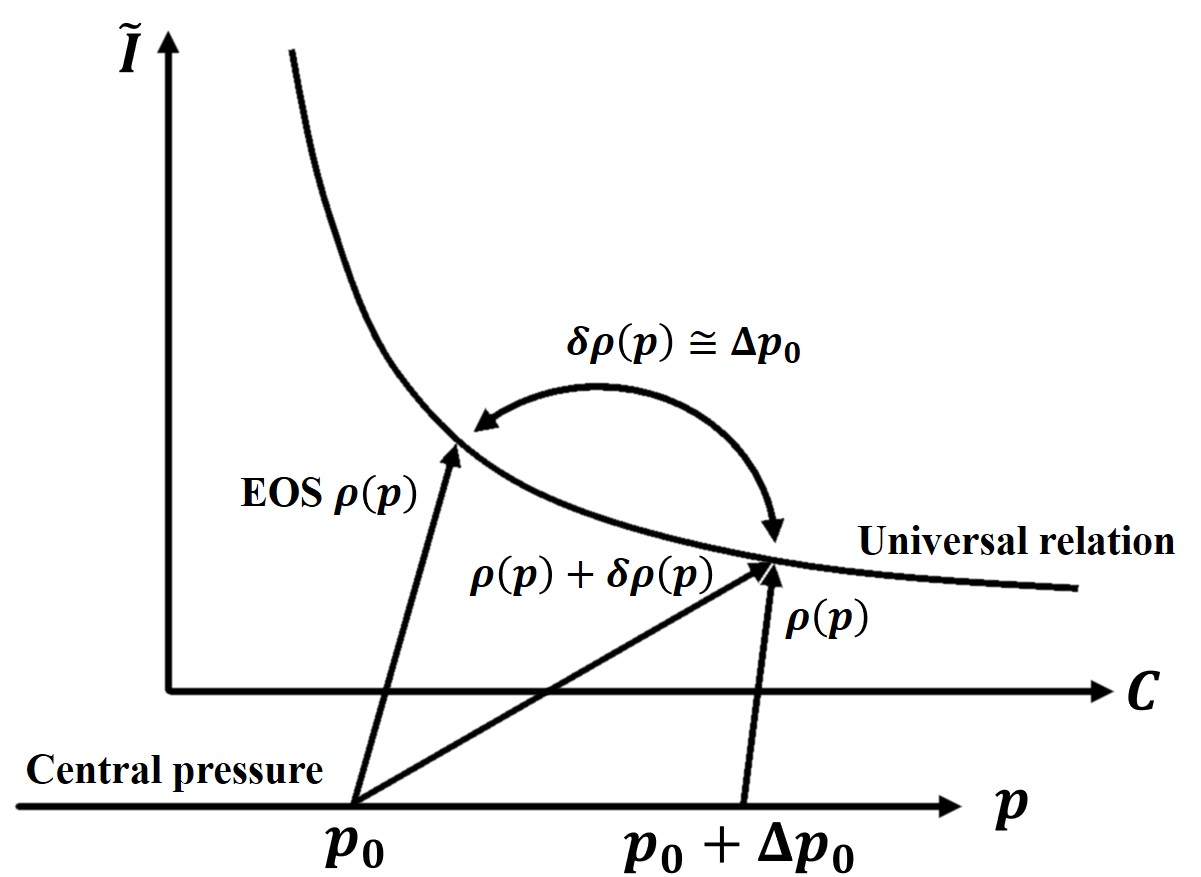}
  \caption{\label{fig:universal_meaning} Illustration of the equivalence 
  between the EOS perturbation and central pressure perturbation for a truly 
  universal relation.}
\end{figure}

It is interesting and intuitive to consider another perturbation here: the 
perturbation of the central pressure. If the relation we consider is truly 
universal, one can conclude that any EOS perturbation should be equivalent to 
some central pressure perturbation with an unperturbed EOS, in the sense that 
they all give points in the unique curve and there is a one to one 
correspondence between the central pressure and the point in this curve. 
We call this as local universality and give an illustration in 
Fig.~\ref{fig:universal_meaning}.

We will consider another star that has EOS $\rho=\rho(p)$ but with the initial
condition
\begin{equation}
  p=p_0+\Delta p\,,\quad 
  r''=0\,,  \quad 
  m''=0\,,\quad 
   i''=0\,.
\end{equation}
Here we use ``$\Delta$'' to distinguish it from the perturbation of EOS.  To
deal with this perturbation, we can integrate one step outwards to get
\begin{equation}\label{eq:icfD}
  p=p_0\,,\quad  
  r''=\Delta r_0\propto |\Delta p|^{1/2}\,, \quad 
   m''=0\,, \quad 
    i''=0\,,
\end{equation}
where we have only kept the leading terms, so that $m''\propto |\Delta p|^{3/2}$
and $i''\propto |\Delta p|^{5/2}$ are ignored. Taking this as the initial
condition, we can write the corresponding solution as
\begin{eqnarray}
  r''(p)&=&r(p)+\Delta r(p)\,,\\
  m''(p)&=&m(p)+\Delta m(p)\,,\\
  i''(p)&=&i(p)+\Delta i(p)\,.
\end{eqnarray}
And again one can derive the equation for the perturbation
\begin{equation}\label{eq:dpc}
  \frac{{\rm d}}{{\rm d}p}
  \begin{bmatrix}
    \Delta r\\
    \Delta m\\
    \Delta i
  \end{bmatrix}
  =\bm{A}(p)
  \begin{bmatrix}
    \Delta r\\
    \Delta m\\
    \Delta i
  \end{bmatrix}\,.
\end{equation}
Note that Eq.~(\ref{eq:dpc}) and Eq.~(\ref{eq:pert}) are exactly the same
equation, but with different initial conditions. Further, as they are linear
equations, scaling the initial condition simply scales the final solution. We
can rescale the initial condition for Eq.~(\ref{eq:pert}) to be
\begin{equation}\label{eq:initial}
  p=p_1\,,\quad 
  \delta r=-r_1\,,\quad 
  \delta m=4\pi r_1^3 p_1\,,\quad 
  \delta i=0\,,
\end{equation}
with $r_1$ the radius where $p=p_1$.  Note that when $p_1\rightarrow p_0$, this
initial condition is just the initial condition in Eq.~(\ref{eq:icfD}). Thus, we
can say that the EOS perturbation at the center of the star is equivalent to the
central pressure perturbation.

The above discussion gives us another method of calculating $K$ in 
Eq.~(\ref{eq:DII_C}), 
\begin{equation}
  K\equiv \frac{{\rm d}\tilde{I}}{{\rm d}C}=\frac{\Delta\tilde{I}}
  {\Delta C}=\frac{\delta\tilde{I}(p_1=p_0)}{\delta C(p_1=p_0)}\,,
\end{equation}
which means that the factor $p_0\rho_0{\rm D}\tilde{I}(p_1)/\tilde{I}$ is always
zero at the center of the star. A necessary condition for the relation to be
truly universal is $\delta\tilde{I}(p_1)/\delta C(p_1)=\Delta \tilde{I}/ \Delta
C$ irrespective of $p_1$, so that the factor $p_0\rho_0{\rm D}\tilde{I}
(p_1)/\tilde{I}$ is always zero throughout the star.

\section{Analytical solution to a constant-density star in the Newtonian
gravity}
\label{sec:example}

It is suggest that the universal relations may come from the similarity of the 
realistic EOSs and an incompressible, constant-density EOS~\cite{Sham:2014kea}. 
Previous study also showed that, the I-Love relation is independent of leading
order EOS perturbation around the incompressible EOS in the Newtonian limit (at
the relative leading order in $C$)~\cite{Chan:2015iou}.  Thus here we study the
I-C relation for a constant-density star in the Newtonian gravity analytically
as an illustration.  

In the Newtonian gravity, after taking a dimensionless form for convenience,
Eq.~(\ref{eq:pert}) is simplified into
\begin{equation}\label{eq:pert_Newt}
  \frac{{\rm d}}{{\rm d}p}
    \begingroup\renewcommand{\arraystretch}{2.5}
  \begin{bmatrix}
    \dfrac{\delta r}{r} \\
    \dfrac{\delta m}{m}\\
    \dfrac{\delta i}{i}
  \end{bmatrix}
  =\begin{bmatrix}
    -\dfrac{r}{\rho m}&\dfrac{r}{\rho m}&0\\
    -\dfrac{16\pi r^4}{m^2}&\dfrac{8\pi r^4}{m^2}&0\\
    -\dfrac{16\pi r^6}{im}&\dfrac{8\pi r^6}{3im}&\dfrac{8\pi r^6}{3im}
  \end{bmatrix}
  \begin{bmatrix}
    \dfrac{\delta r}{r}\\
    \dfrac{\delta m}{m}\\
    \dfrac{\delta i}{i}
  \end{bmatrix}\,,
  \endgroup
\end{equation}
with initial conditions
\begin{equation}\label{eq:Newtonian_ini}
  p=p_1\,,\quad 
  \delta r/r=-r_1/m_1\rho_1^2\,,\quad 
  \delta m/m=0\,,\quad 
  \delta i/i=0\,,
\end{equation}
where $r_1=r(p_1)$ and $m_1=m(p_1)$. For a constant-density star, one has
\begin{equation}
  4\pi r^3\rho/m=3\,,\quad 
  8\pi r^5\rho/i=15\,.
\end{equation}
Thus, by introducing the following combination,
\begin{eqnarray}
  \delta u&=&\frac{\delta r}{r}-\frac{1}{4}\frac{\delta m}{m}\,,\\
  \delta v&=&\frac{\delta r}{r}-\frac{1}{3}\frac{\delta m}{m}\,,\\
  \delta w&=&\frac{\delta i}{i}-20\delta u+15\delta v\,,
\end{eqnarray}
Eq.~(\ref{eq:pert_Newt}) can be directly solved as
\begin{subequations}\label{eq:const_sol}
  \begin{eqnarray}
    \delta u(p;p_1)&=&-\frac{r_1}{m_1\rho_1^2}e^{\int_{p_1}^p\frac{2r}
    {\rho m}{\rm d}p}\,,\\
    \delta v(p;p_1)&=&-\frac{r_1}{m_1\rho_1^2}
    e^{\int_{p_1}^p\frac{3r}{\rho m}{\rm d}p}\,,\\
    \delta w(p;p_1)&=&\frac{5r_1}{m_1\rho_1^2} e^{\int_{p_1}^p\frac{5r}{\rho m}{\rm d}p}\,.
  \end{eqnarray}
\end{subequations}

Note that the integral
\begin{equation}
  \int_{p_1}^p\frac{r}{\rho m}{\rm d}p=\int_{r}^{r_1}\frac{{\rm d}r}{r}=\ln\frac{r_1}{r}\,,
\end{equation}
for a constant-density star. Combining with Eq.~(\ref{eq:const_sol}), we have 
\begin{eqnarray}\label{eq:DII_ingre}
  \frac{{\rm D}\tilde{I}(p_1)}{\tilde{I}}&=&\frac{\delta I(p_1)}{I}-
  3\frac{\delta M(p_1)}
  {M}-K\frac{C}{\tilde{I}}\left(\frac{\delta M(p_1)}{M}-\frac{\delta R(p_1)}
  {R}\right)\nonumber\\
  &=&\delta w(p=0;p_1)+3\delta v(p=0;p_1)\nonumber\\
  &=&-\frac{3r_1}{m_1\rho^2}\left(\frac{r_1}{R}\right)^3+
  \frac{5r_1}{m_1\rho^2}\left(\frac{r_1}{R}\right)^5\,,
\end{eqnarray}
where we have used $K=-2\tilde{I}/C$ for a constant-density star in the
Newtonian gravity.  Taking into account that 
\begin{eqnarray}
  m(r)&=&\frac{4\pi}{3}r^3\rho\,,\\
  r(p)&=&\sqrt{\frac{3}{2\pi\rho^2}}(p_0-p)^{1/2}\,,
\end{eqnarray}
we have
\begin{equation}\label{eq:const_fac1}
  p_0\rho_0\frac{{\rm D}\tilde{I}(p_1)}{\tilde{I}}=\left(1-\frac{p_1}{p_0}\right)^{1/2}
  \left(1-\frac{5}{2}\frac{p_1}{p_0}\right)\,.
\end{equation}
Interestingly, the above factor only depends on the ratio $p_1/p_0$ thus is 
universal for all background stars. As we will see later, for realistic EOSs,
this factor in general also depends on the compactness of the background star.
Though it is in the Newtonian gravity and for  a constant-density star,
Eq.~(\ref{eq:const_fac1}) still captures the order of magnitude of  this factor,
thus provides a very intuitive estimation of the universality of the I-C
relation.

\section{Linear response for compact stars with realistic 
EOSs}\label{sec:real}

\begin{figure}[t]
  \centering
  \includegraphics[width=0.45\textwidth]{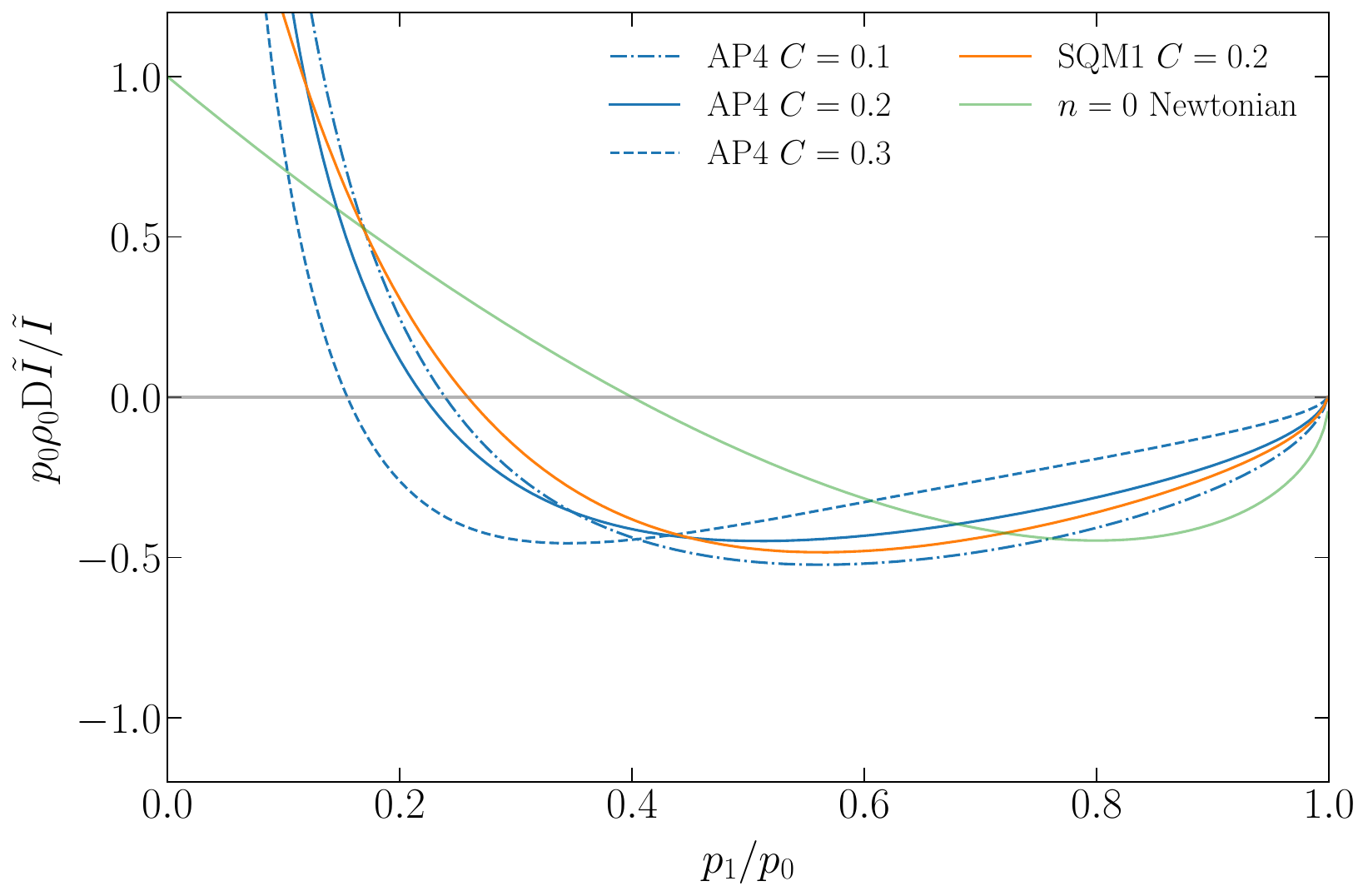}
  \caption{\label{fig:fac1} The factor $p_0\rho_0 {\rm D}\tilde{I}/\tilde{I}$ as
  a function of $p_1/p_0$. Note that unlike the analytical case we discussed
  before, for realistic EOSs, the factor in general depends on the background
  stars.  Here we show the factor for NSs with the AP4 EOS and different
  compactnesses as well as for QSs with the SQM1 EOS. We also show this factor
  for a polytropic EOS with $n=0$ in the Newtonian gravity given in
  Eq.~(\ref{eq:const_fac1}). }
\end{figure}

In this section, we focus on NSs and QSs with realistic EOSs in GR. We use the 
EOS data provided by~\citet{Lattimer:2000nx}, which are in a tabulated form. The EOSs for 
NS have a crust EOS down to a density of $10^8\,{\rm g\,cm^{-3}}$. 
To obtain
the factor $p_0\rho_0{\rm D}\tilde{I}(p_1)/\tilde{I}$ for realistic EOSs, in
general one needs to numerically solve Eq.~(\ref{eq:pert}). In 
Fig.~\ref{fig:fac1}, we show this factor for NSs with the AP4 
EOS~\cite{Akmal:1998cf} as well as QS with the SQM1 EOS~\cite{Prakash:1995uw}. 
As in general this factor depends on the background stars, we plot the NS cases 
for three different star compactnesses. We also plot Eq.~(\ref{eq:const_fac1}) 
for comparison. One can see that, the Newtonian constant density approximation 
does capture the order of magnitude for the factor in realistic cases, except 
for a small region where $p_1/p_0$ is close to zero, that is, the outer region
of the NS. We find that when $p_1/p_0$ tends to zero, this factor grows up 
towards infinity (or towards a relatively large but finite number in the case of
QSs) for realistic EOSs. We will discuss this behavior later in detail as it is
closely related to the fact that NSs and QSs have different branches of the I-C
relation.  

\begin{figure}[t]
  \centering
  \includegraphics[width=0.45\textwidth]{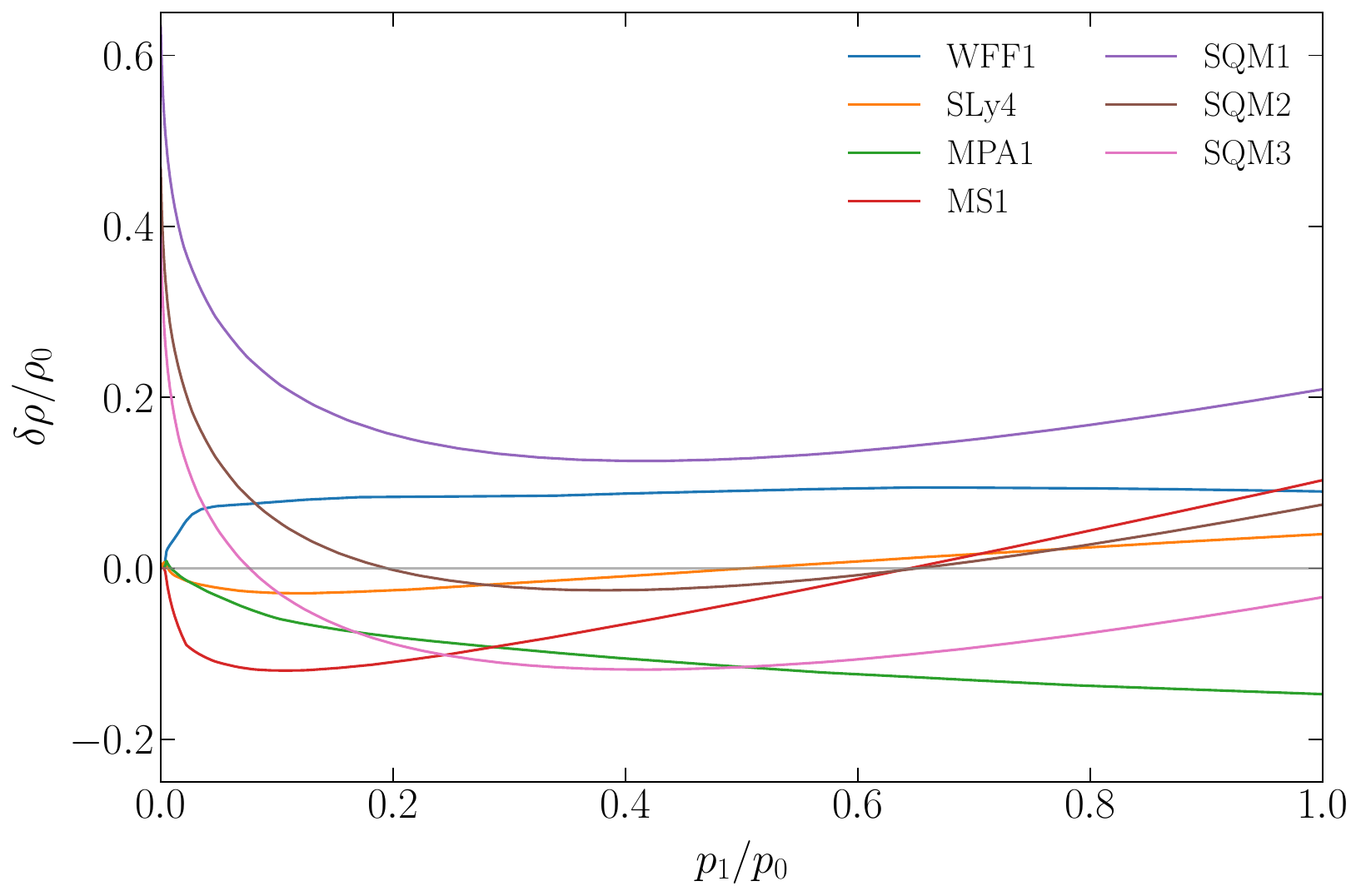}
  \caption{\label{fig:fac2} The factor $\delta\rho/\rho_0$ for different 
  realistic EOSs relative to the AP4 EOS. The background star has a compactness 
  $C=0.2$. Different from the NS-NS case, the difference between the QS EOS and
  the NS EOS is finite at $p_1/p_0=0$.}
\end{figure}
\begin{figure}[t]
  \centering
  \includegraphics[width=0.45\textwidth]{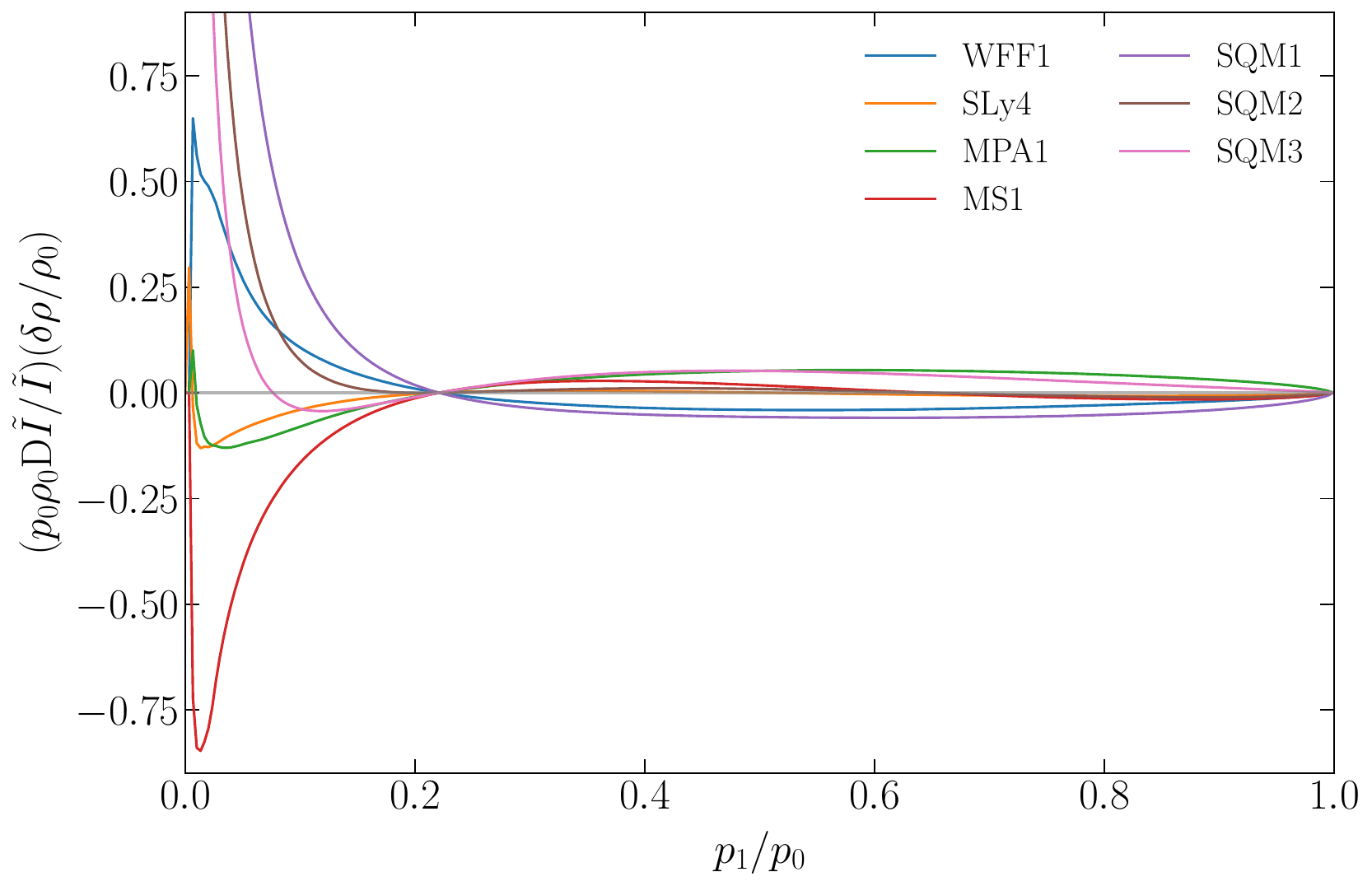}
  \caption{\label{fig:product} The product of the two factors from
  Fig.~\ref{fig:fac1} and Fig.~\ref{fig:fac2}. The relative deviation then can
  be ``read off'' from the figure as the area enclosed by the curve and the
  $x$-axis.}
\end{figure}
\begin{figure}[t]
  \centering
  \includegraphics[width=0.45\textwidth]{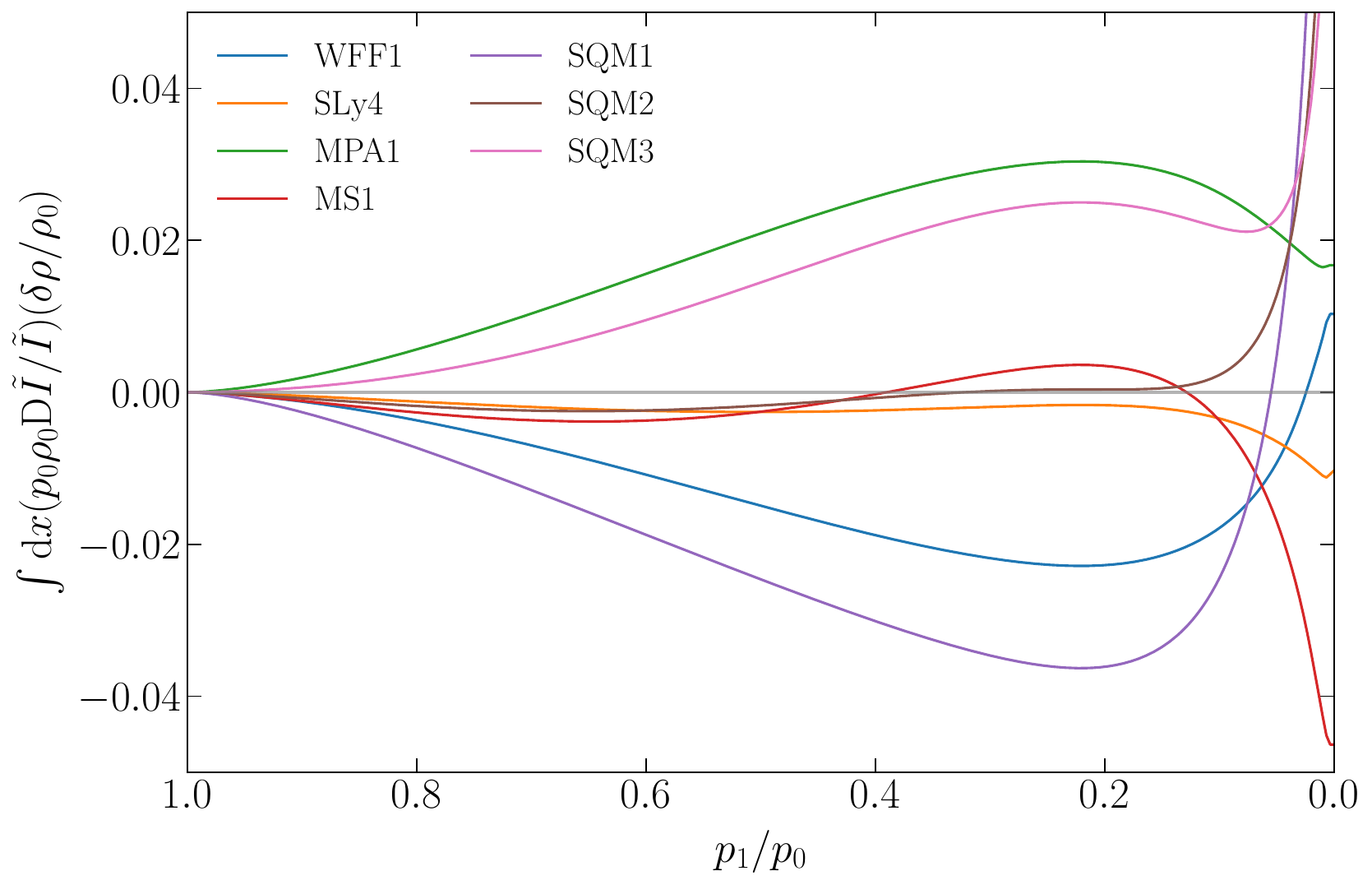}
  \caption{\label{fig:Int_product} The integration cumulated from the center of
  the star to its surface.}
\end{figure}

To estimate the integral in Eq.~(\ref{eq:pert_integral}), which represents the 
deviation of the relation caused by the difference between realistic EOSs, in 
Fig.~\ref{fig:fac2} and Fig.~\ref{fig:product}, we show the factor
$\delta\rho(x)/\rho_0$ and the product of the two factors for some realistic EOS
perturbations, respectively. Here $\delta \rho$ represents the difference 
between the AP4 EOS and the other EOSs. From the figures, one can easily
estimate that  the universal relation will hold at  a percent level. The sign 
change further minishes the integral. One can also roughly conclude that the
mean contribution of the violation of the relation comes from the outer region
of the star. This is  shown in Fig.~\ref{fig:Int_product}, where we plot the
integration cumulated from the center of the star where $p_1/p_0=1$ to its
surface where $p_1/p_0=0$.  It is clear that, when considering the deviation
caused by the EOS difference between a QS EOS and a background NS EOS, the
integral becomes divergent at the surface of the star, which is caused by the
finite density difference at $p=0$ for a NS and a QS as well as the divergent
behavior of the factor $p_0\rho_0{\rm D}\tilde{I}(p_1)/\tilde{I}$. 

\begin{figure}[t]
  \centering
  \includegraphics[width=0.45\textwidth]{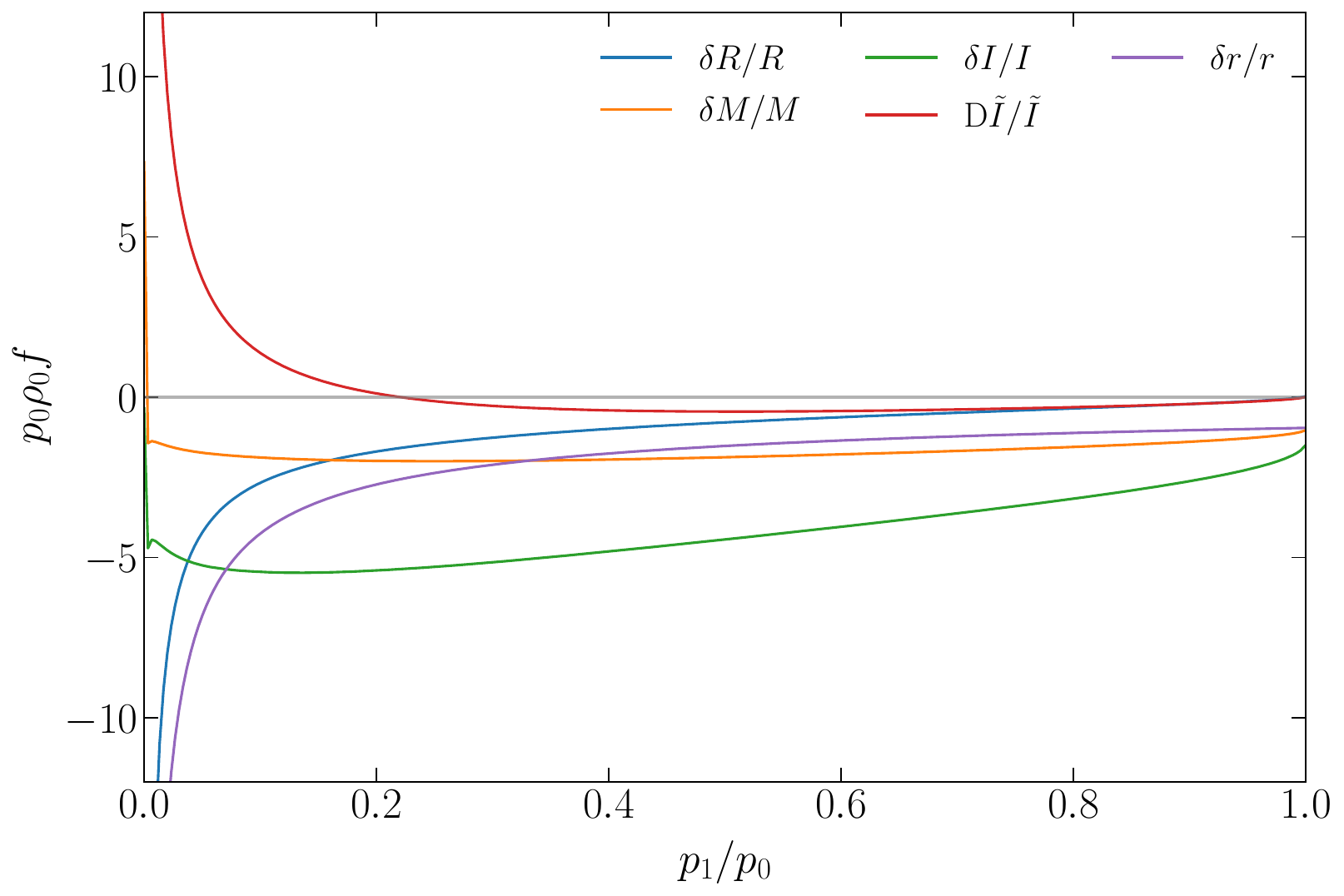}
  \caption{\label{fig:fac_behavior} The behavior of different factors that
  compose ${\rm D}\tilde{I}/\tilde{I}$  for the AP4 EOS and a star compactness
  $C=0.2$. $f$ in the $y$-axis label represents for $\delta R/R$, $\delta M/M$
  and so on. The divergent behavior comes from $\delta R/R$ and can be
  understood with the initial condition $\delta r /r$.}
\end{figure}

To understand this divergent behavior, we notice that in 
Eq.~(\ref{eq:DII_ingre}), the behavior of the factor $p_0\rho_0{\rm
D}\tilde{I}(p_1)/\tilde{I}$ depends on $\delta I(p_1)/I$, $\delta M(p_1)/M$, and
$\delta R(p_1)/R$, which are directly solved from the perturbation equations. 
In Fig.~\ref{fig:fac_behavior}, we plot these factors separately. From this
figure, one can clear see that the divergent behavior at the surface of the star
is dominated by the factor $\delta R(p_1)/R$. 

The divergence of $\delta R(p_1)$ near the surface of the NS can be analytically 
estimated via a polytropic crust EOS model. Considering EOS perturbations in the low 
pressure region, the inner structure of the star then is irrelevant to our problem 
and the effect can be fully characterized by the statement that, at $p=p_R$ and 
$\rho=\rho_R$, the core of the star has mass $M$, radius $R$, where $p_R$ is some 
small pressure. We can then assume that for $p<p_R$, the NS crust has a polytropic 
EOS that $\rho(p)=\rho_R (p/p_R)^{n/(n+1)}$ and we can model the crust of the NS with 
Newtonian approximation,
\begin{equation}
  \frac{{\rm d}r}{{\rm d}p}=-\frac{1}{\rho}\frac{r^2}{m}\,.
\end{equation}
Further, it is known that the crust of NS contains only a small 
amount of mass but provides a relatively considerable radius~\cite{Chan:2014tva}, therefore, 
when considering the behavior of $\delta r$, we can set $m=M$ in the above equation (also note that 
$\delta r$ is the only divergent part so we can safely ignore $\delta m$ there).

With these approximation, one can solve the structure of the curst analytically, which reads
\begin{equation}
  \frac{1}{r}-\frac{1}{R}=(n+1)\frac{p_R^{\frac{n}{n+1}}}{M\rho_R}\left(p^{\frac{1}{n+1}}-p_R^{\frac{1}{n+1}}\right)\,.
\end{equation} 
This gives a crust with a depth $z\approx(n+1)R^2p_R/M\rho_R$. Following the procedure introduced in 
Sec.~\ref{sec:local U}, one can then write down the equation for $\delta r(p)$ caused by an EOS perturbation 
$\delta \rho(p)=\delta(p-p_1)$ as 
\begin{equation}
  \frac{{\rm d}\delta r}{{\rm d}p}=-\frac{2r}{\rho M}\delta r\,,
\end{equation}
with initial condition $\delta r(p_1)=-r(p_1)^2/\rho(p_1)^2M$. This equation again can be integrated 
analytically and one can obtain
\begin{equation}
  \delta R(p_1)=\delta r(p_1)\left(\frac{\frac{M\rho_R/R-p_R}{p_R^{n/(n+1)}}+p_1^{1/(n+1)}}
    {\frac{M\rho_R/R-p_R}{p_R^{n/(n+1)}}}\right)^2\,\approx-\frac{R^2}{M\rho(p_1)^{2}}.
\end{equation}
Thus, near the surface $\delta R(p_1)/R$ diverges as $\rho(p_1)^{-2}$. In Fig.~\ref{fig:fac_behavior}, we 
also plot the above estimation with $M/R\approx 0.2$. It can be seen that this estimation captures the 
divergent behavior of $\delta R(p_1)/R$. Differently for QSs, near the surface, they behave more like 
the constant density star as we discussed before. Factors remain finite there.

We should note that, the divergence of $\delta R(p_1)$ is a physical result and itself does not cause 
problem to our linear analysis. The quantity that needs to be finite and small is the integral defined 
in Eq.~(\ref{eq:inte_R}) when the linear approximation holds. As an example, we may consider the EOS 
perturbation with $n\rightarrow n+\epsilon$ in the above crust model. The corresponding change in the 
density is 
\begin{equation}
  \delta \rho(p)=\frac{\epsilon\rho_R}{(n+1)^2}\left(\frac{p}{p_R}\right)^{\frac{n}{n+1}}\ln\frac{p}{p_R}\,,
\end{equation}
which is well defined for $n>0$. The integration, 
\begin{equation}
  \int_0^{p_R}{\rm d}p_1\delta R(p_1)\delta\rho(p_1)\approx \epsilon \frac{R^2p_R}{M\rho_R}\,,
\end{equation}
is just the change in $z$ that $\delta z=\delta n R^2 p_R/M\rho_R=\epsilon R^2p_R/M\rho_R$. One can 
also notice that, in Fig.~\ref{fig:Int_product}, the integration of ${\rm D}\tilde{I}$ is finite for 
NS-NS comparison. As $\delta R(p_1)$ is the only divergent part in ${\rm D}\tilde{I}$, the integration 
of $\delta R(p_1)$ for realistic NS EOSs, remains finite.

\begin{figure}[t]
  \centering
  \includegraphics[width=0.45\textwidth]{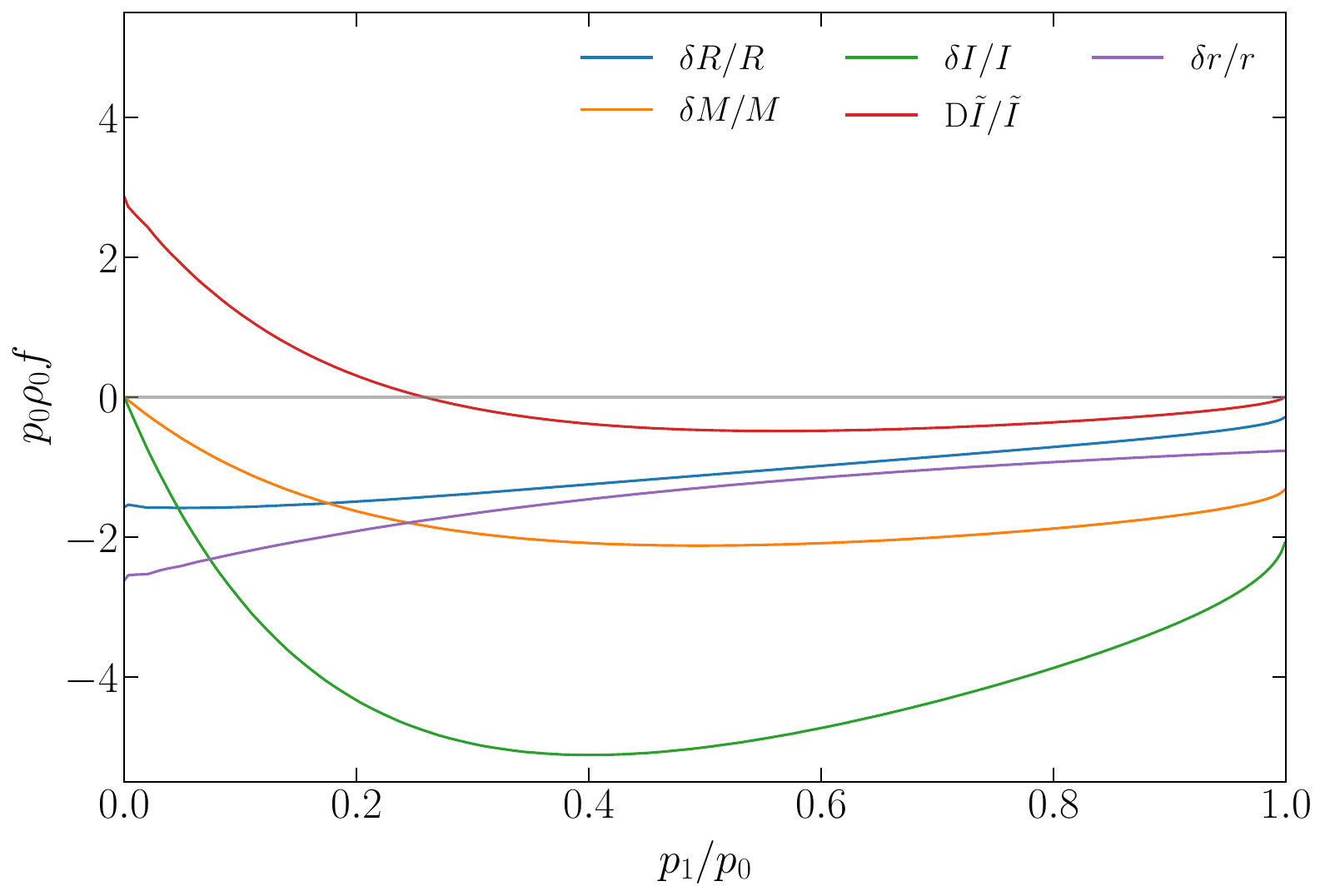}
  \caption{\label{fig:fac_behavior_QS} The behavior of different factors that
  compose ${\rm D}\tilde{I}/\tilde{I}$ for the SQM1 EOS and a star compactness
  $C=0.2$.}
\end{figure}

Finally, the integrated factor $p_0\rho_0{\rm D}\tilde{I}/\tilde{I}$ is
divergent if one considers a NS background but takes a QS EOS as the perturbed
EOS. As the 
QS EOS has a finite surface density $\rho_s$ while NS EOS at the low density part 
can be approximated by $\rho\propto p^{3/5}$, the integration at small $p_1$ 
is roughly $\int 1/\rho_1^2 {\rm d}p_1$ and divergent. As we will
see in Sec.~\ref{sec:linear}, this divergence means that the linear approximation fails for this
situation, suggesting an intrinsic difference in the I-C relation for NSs and
QSs. We also 
note that, the equivalence between taking any one of the star as background only 
holds when both choices give finite results and it does not apply to the NS-QS comparison.

\section{I-Love relation}\label{sec:I-Love}

It is known that the I-Love universal relation holds better compared to the I-C 
relation. Previous study showed that the leading order dependence on $C$ cancels
out for perturbation of the I-Love relation~\cite{Chan:2015iou}. In this
section, we apply our approach introduced in previous sections to the I-Love
relation.

\begin{figure}[t]
  \centering
  \includegraphics[width=0.45\textwidth]{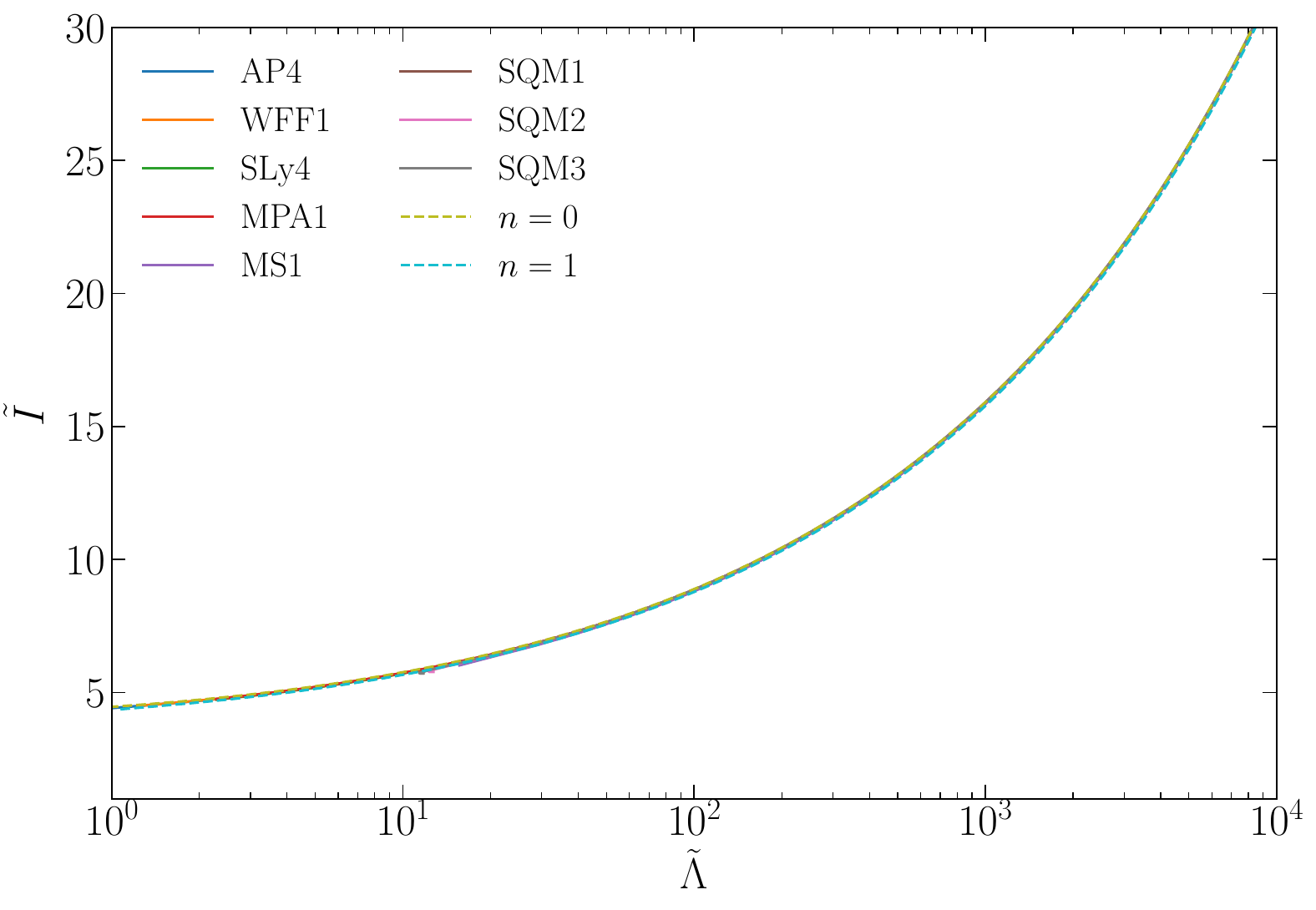}
  \caption{\label{fig:universal_I_L} The I-Love relation for NSs and QSs with
  realistic EOSs as well as polytropic EOSs. The dimensionless moment of
  inertial is defined to be $\tilde{I}\equiv I/M^3$ and the dimensionless tidal
  deformability $\tilde{\Lambda} \equiv \Lambda/M^5$. 
  }
\end{figure}

Tidal love number in GR can be calculated following the procedure developed by
\citet{Hinderer:2007mb}, and here we  list the relevant results for convenience.
The  tidal deformability is,
\begin{equation}\label{eq:k2L}
  \Lambda=\frac{2}{3}k_2R^5\,,
\end{equation}
where
\begin{eqnarray}\label{eq:Y2k}
  k_2&=&\frac{8C^5}{5}(1-2C)^2 \big[ 2+2C(Y-1)-Y \big]\nonumber\\
  &\times& \Big\{2C \big[ 6-3Y+3C(5Y-8) \big] \nonumber\\
  &+&4C^3 \big[13-11Y+C(3Y-2)+2C^2(1+Y) \big]\nonumber\\
  &+&3(1-2C)^2 \big[2-Y+2C(Y-1) \big]\ln(1-2C) \Big\}^{-1}\,,
\end{eqnarray}
is the dimensionless Love number; $Y = y \big|_{r=R} $ needs to be calculated
from the ordinary differential equation
\begin{eqnarray}
  \frac{{\rm d}y}{{\rm d}r}&=&\left(\frac{1}{r}-A\right)y-\frac{y^2}{r}-rB\,,
\end{eqnarray}
where 
\begin{eqnarray}
  A&=&\frac{2}{r}+\frac{1}{1-2m/r}\left[\frac{2m}{r}+4\pi r(p-\rho)\right]\,,\\
  B&=&-\frac{6}{r^2(1-2m/r)}+\frac{4\pi}{1-2m/r}\Bigg[5\rho+9p \nonumber\\
  &+& (\rho+p)\frac{{\rm d}\rho}{{\rm d}p}\Bigg]-\left(2\frac{m+4\pi r^3 p}
  {r^2(1-2m/r)}\right)^2\,,
\end{eqnarray}
with an initial value $y\big|_{r=0}=2$.  In Fig.~\ref{fig:universal_I_L} we show
the I-Love relation for several realistic EOSs as well as polytropic EOSs. This
universal relation holds better compared to the I-C relation at $\sim 0.1\%$
level. This relation is also universal among NSs and QSs.

By adding an EOS perturbation with a form $\delta \rho=\delta(p-p_1)$ and
solving the perturbation quantities to the leading order, one can obtain a
similar factor $\delta \Lambda/ \Lambda$. The concerned quantity now is 
\begin{equation}\label{eq:DII_L}
  \left(\frac{{D}\tilde{I}}{\tilde{I}}\right)_{\tilde{\Lambda}}\equiv
  \frac{\delta \tilde{I}-K_{\tilde{\Lambda}}\delta \tilde{\Lambda}}{\tilde{I}}\,,
\end{equation}
where we use the subindex $\tilde{\Lambda}$ to distinguish it from the quantity
before. However, due to the term ${\rm d}\rho/{\rm dp}$ in the differential
equation, additional care needs to be taken for calculating $\delta
\Lambda/\Lambda$. In Appendix~\ref{app:I-L}, we give the detailed calculation
procedure for this factor.

Though for the general cases in GR, the numerical calculation takes some
efforts, it is still easy and intuitive to have an analytical solution for this
factor in the Newtonian gravity for a constant-density star as we do for the I-C
relation. Remind that in the Newtonian gravity, the tidal Love number is
governed by the Clairaut-Radau equation~\cite{poisson_will_2014},
\begin{equation}\label{eq:CReq}
  r\frac{{\rm d}\eta_l}{{\rm d}r}+\eta_l(\eta_l-1)+6\frac{4\pi
  r^3\rho}{3m}(\eta_l+1) -l(l+1)=0\,,
\end{equation}
with an initial condition $\eta_l \big|_{r=0}=l-2$. The dimensionless Love
number is given by 
\begin{equation}
  k_l=\frac{l+1-\eta_l(R)}{2 \big[l+\eta_l(R) \big]}\,.
\end{equation}

For a constant-density star and the $l=2$ Love number that we are interested,
the background solution is simply $\eta_2(r)=0$ and $k_2=3/4$. Perturbing
Eq.~(\ref{eq:CReq}) with $\delta \rho(p)=\delta (p-p_1)$ and simplifying with
the background solution, one gets 
\begin{equation}
  \frac{{d}\delta \eta}{{\rm d}p}=\frac{18r}{\rho m}\frac{{\rm
  d}r}{r}-\frac{6r}{\rho m} \frac{\delta m}{m}+\frac{5r}{\rho m}\delta \eta\,,
\end{equation} 
with initial conditions 
\begin{equation}
  p=p_1\,,\quad 
  \delta \eta=-\frac{6r_1}{m_1 \rho_1^2}\,.
\end{equation}
The combination 
\begin{equation}
  \delta z=\delta \eta +9\frac{\delta r}{r}-3\frac{\delta m}{m}\,,
\end{equation}
then solves the equation with 
\begin{equation}
  \delta z(p;p_1)=-\frac{15 r_1}{m_1\rho_1^2}e^{\int_{p_1}^p\frac{5r}{\rho
  m}{\rm d}p}\,.
\end{equation}
Using the fact that 
\begin{equation}
  \frac{\delta \Lambda}{\Lambda}=\frac{\delta k_2}{k_2}+5\frac{\delta R}{R}
  =-\frac{5}{6}\delta \eta+5\frac{\delta R}{R}\,,
\end{equation}
and $K_{\tilde{\Lambda}}=2/5$ for a constant-density star in the Newtonian
gravity and taking into account the solutions in Eq.~(\ref{eq:const_sol}), one
can easily verify that, the factor defined in Eq.~(\ref{eq:DII_L}) is
\begin{equation}
  \left(\frac{{\rm D}\tilde{I}(p_1)}{\tilde{I}}\right)_{\tilde{\Lambda}}=0\,.
\end{equation}
This result is in fact equivalent to the result obtained by
\citet{Chan:2015iou}, where they expanded the EOS with $\rho (p)=c_0+c_1 p+c_2
p^2+\cdots$ and expanded the star structure with $C$. They found that the I-Love
relation does not depend linearly in $c_n$ to the leading order of $C$ for
$n\neq 0$. As the polynomial expansion of EOSs does span all the possible 
perturbations and the leading order of $C$ is equivalent to the Newtonian limit,
while linear in $c_n$ is linear in the EOS perturbation, their result has
exactly the same meaning as we obtain here. However, we get this result with a 
different approach that is more convenient to consider arbitrary realistic EOS
perturbations and can be simply extended to full GR cases.

\begin{figure}[t]
  \centering
  \includegraphics[width=0.45\textwidth]{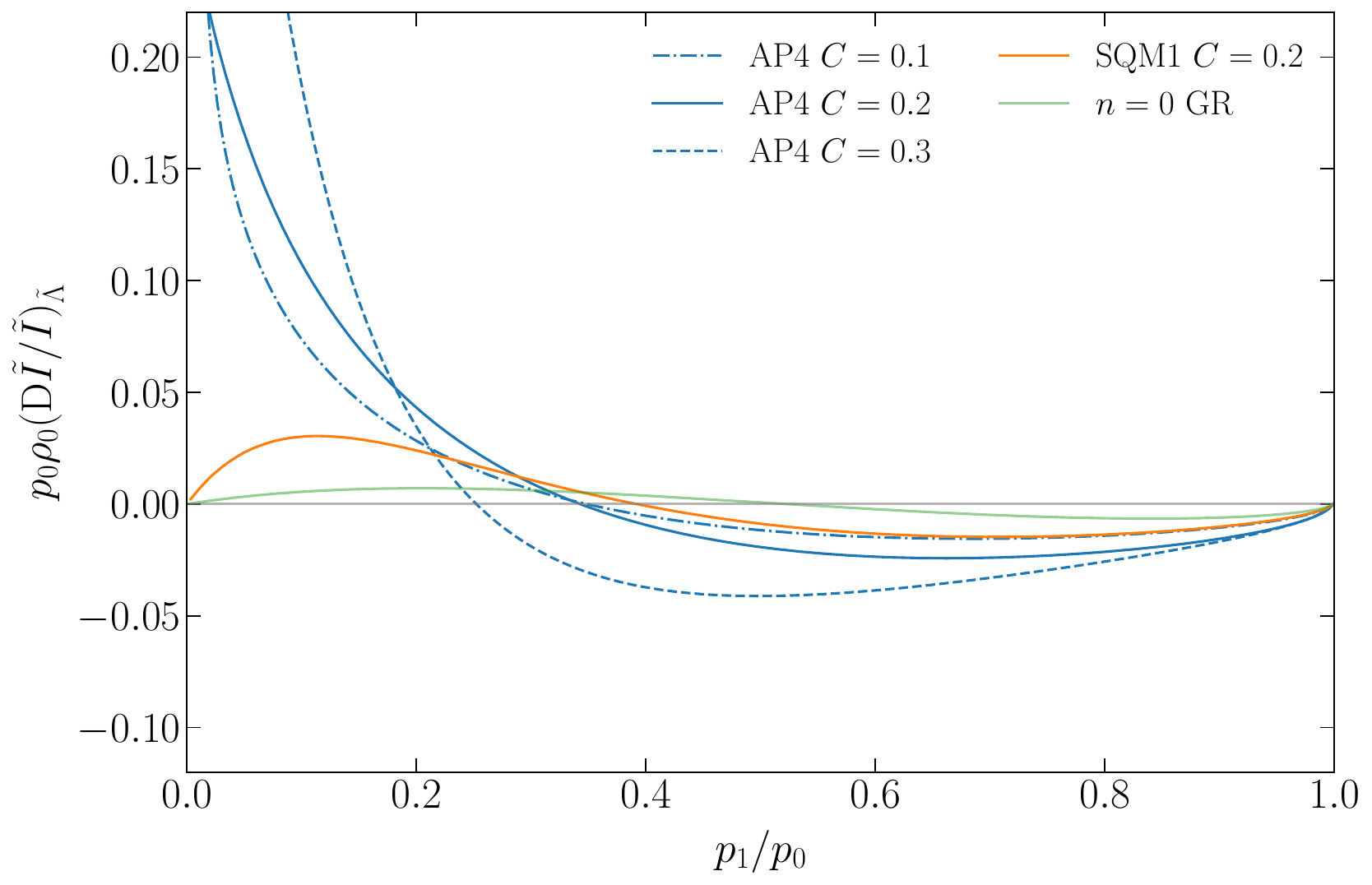}
  \caption{\label{fig:fac1_L} The factor $p_0\rho_0 ({\rm D}\tilde{I}/\tilde{I})
  _{\tilde{\Lambda}}$ as a function of $p_1/p_0$. As before, we plot this factor
  for the AP4 and SQM1 EOSs. Here we also plot the factor for the
  constant-density star in GR with $C=0.2$, while its Newtonian counterpart
  leads to zero. This factor is similar to the factor for the I-C relation but
  is an order of magnitude smaller, providing a better universality in the
  I-Love relation.}
\end{figure}

Figure~\ref{fig:fac1_L} shows the factor $p_0\rho_0({\rm D}\tilde{I}/\tilde{I})_
{\tilde{\Lambda}}$ for several different background stars in GR. It can be seen
that, compared to the factor for the I-C relation, this factor is about an order
of magnitude smaller, thus provides a much better universality. Different from
the factor for the I-C relation, here for QSs, at the surface of the star the
factor seems approaching zero, leading to a weaker response to the EOS
perturbation, which is consistent with the suggestion that the I-Love relation
comes from the constant-density property~\cite{Sham:2014kea}.  We also find
that, for a polytropic EOS with $n=0$ in GR and $C=0.2$, this factor is 
relatively small but not strictly zero. However, this smallness may not be able
to be simply attributed as a continuation from the Newtonian limit. $C=0.2$ is 
already a large compactness if one notices that, the second order term in $C$
for the expansion of $\tilde{\Lambda}$ given in Ref.~\cite{Chan:2015iou},
Eq.~(6.28), is already larger than the leading order term in this case.

\section{Validity of the linear order analysis}\label{sec:linear}

\begin{figure}[t]
  \centering
  \includegraphics[width=0.45\textwidth]{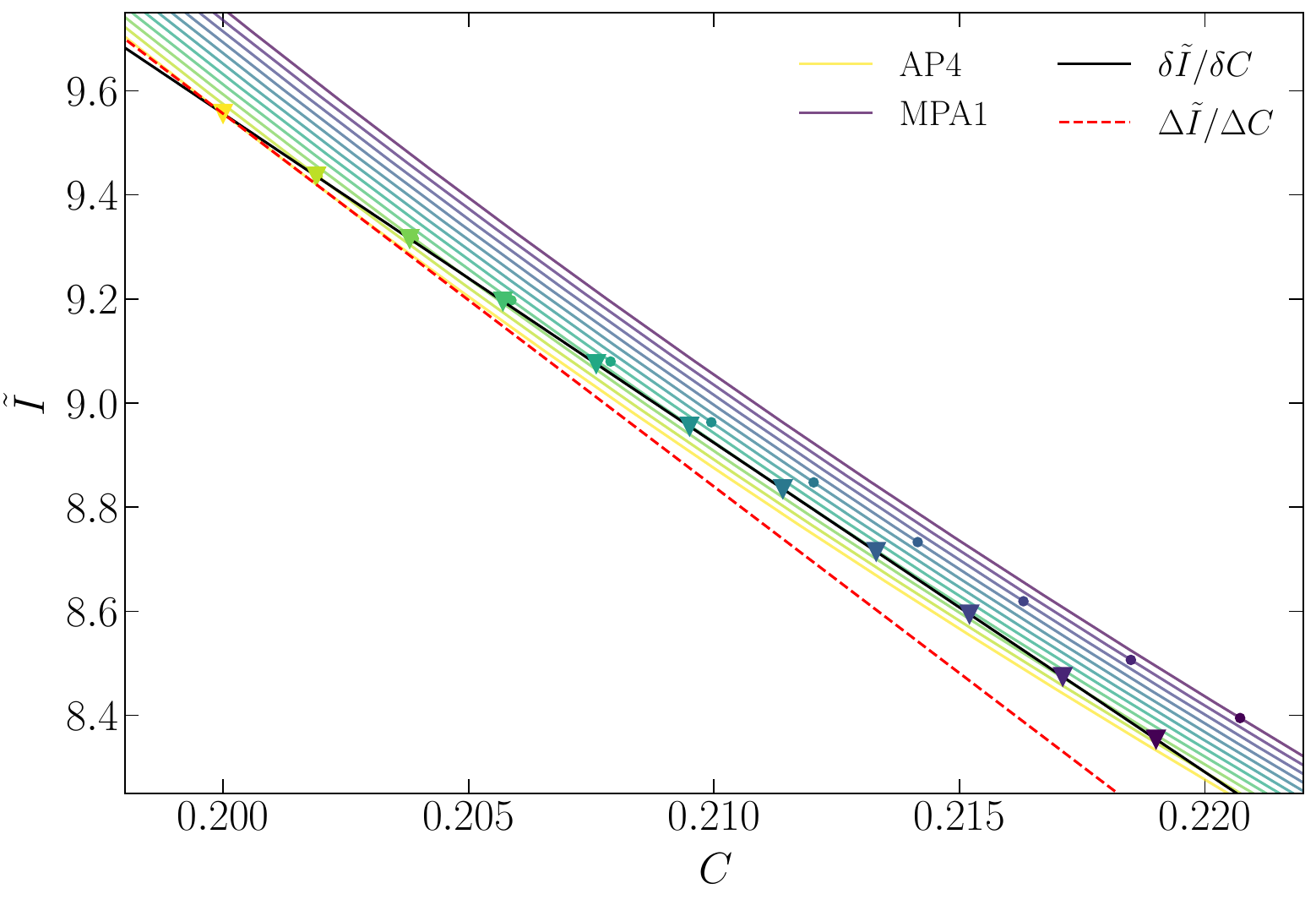}
  \caption{\label{fig:I_C_11} The linear perturbation prediction for the I-C
  relation for EOS perturbations that continuously connect the AP4 EOS and the
  MPA1 EOS. The black line denotes the linear perturbation prediction while the
  red dashed line is the tangent line at the background point. The circles mark
  the star corresponding to the same central pressure as the background star for
  each EOS, while the triangles mark the corresponding linear prediction.}
\end{figure}
\begin{figure}[t]
  \centering
  \includegraphics[width=0.45\textwidth]{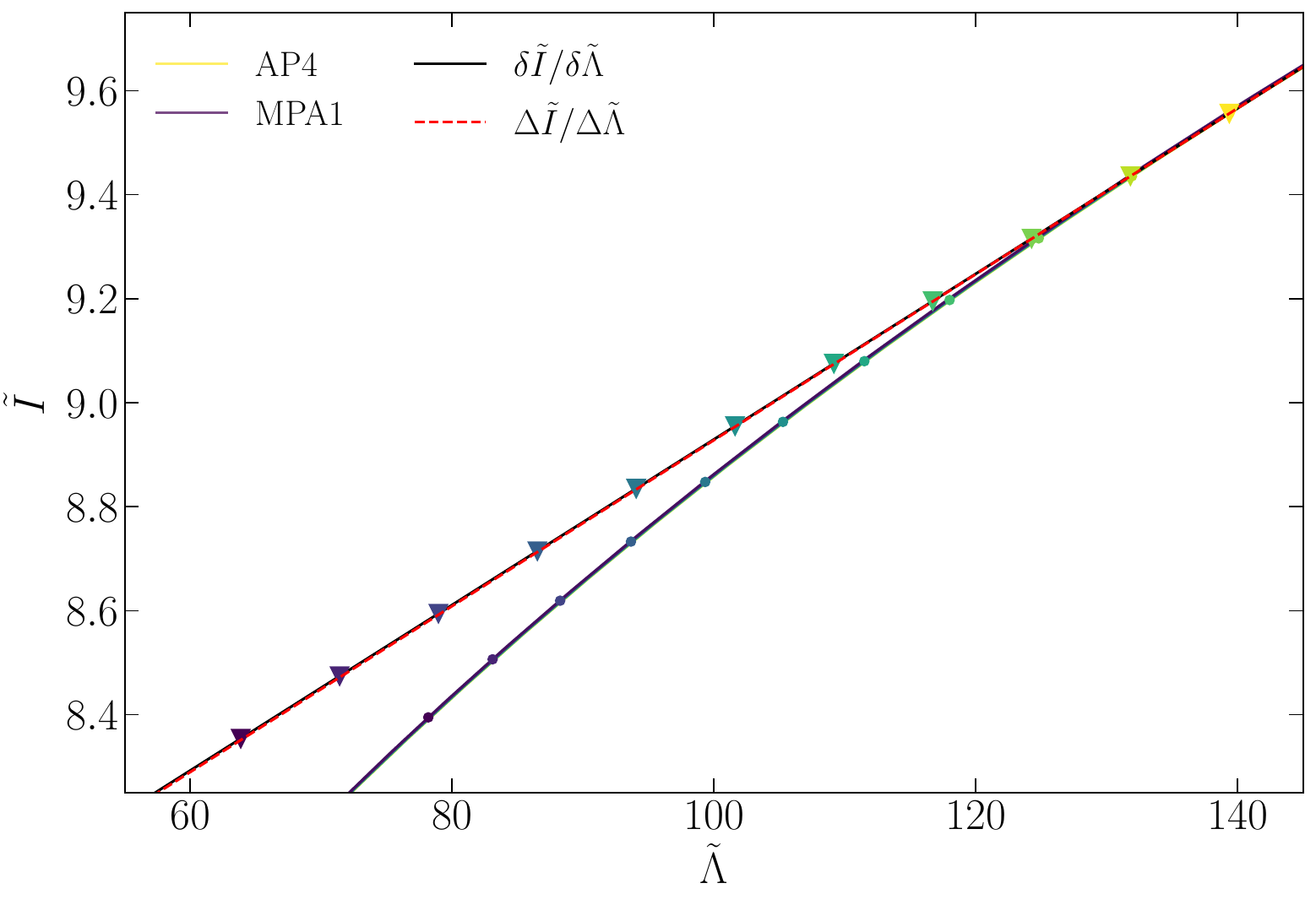}
  \caption{\label{fig:I_L_11} Similar to Fig.~\ref{fig:I_C_11}, but for the 
  I-Love relation.}
\end{figure}

In previous sections, we discuss the linear response of the I-C and I-Love
universal relations under EOS perturbations. One natural question is that to
what extend this linear approximation holds and whether it can be applied to
realistic EOSs.  In this section, we discuss this question based on examples of
realistic EOS differences.

Taking AP4 and MPA1 as example, in Fig.~\ref{fig:I_C_11} and
Fig.~\ref{fig:I_L_11}, we plot the linear perturbation prediction for these two
EOSs. Defining 
\begin{equation}
  \delta\rho(p)=\rho_{\rm MPA1}(p)-\rho_{\rm AP4}(p)\,,
\end{equation}
we further construct a series of 11 EOSs that connect the AP4 EOS and the MPA1
EOS via 
\begin{equation}
  \rho_{i}(p)=\rho_{\rm AP4}(p)+i\delta\rho(p)/10\,,\quad i=0,\cdots,10\,.
\end{equation}
In Fig.~\ref{fig:I_C_11} and Fig.~\ref{fig:I_L_11}, we plot the relation for all
these EOSs. Taking the star with  a compactness $C=0.2$ and the AP4 EOS as the
background star, we also denote the stars with the same central pressure in
different EOSs. We can see that, the linear theory predicts the perturbation
well for small EOS deviations, however, the real difference between the AP4 EOS
and the MPA1 EOS seems to be out of the linear range. We also note that, if the
relation is truly universal, the black line should coincide with the red dashed
line. As expected, the I-Love relation shows better universality at the linear 
order.

\begin{figure}[t]
  \centering
  \includegraphics[width=0.45\textwidth]{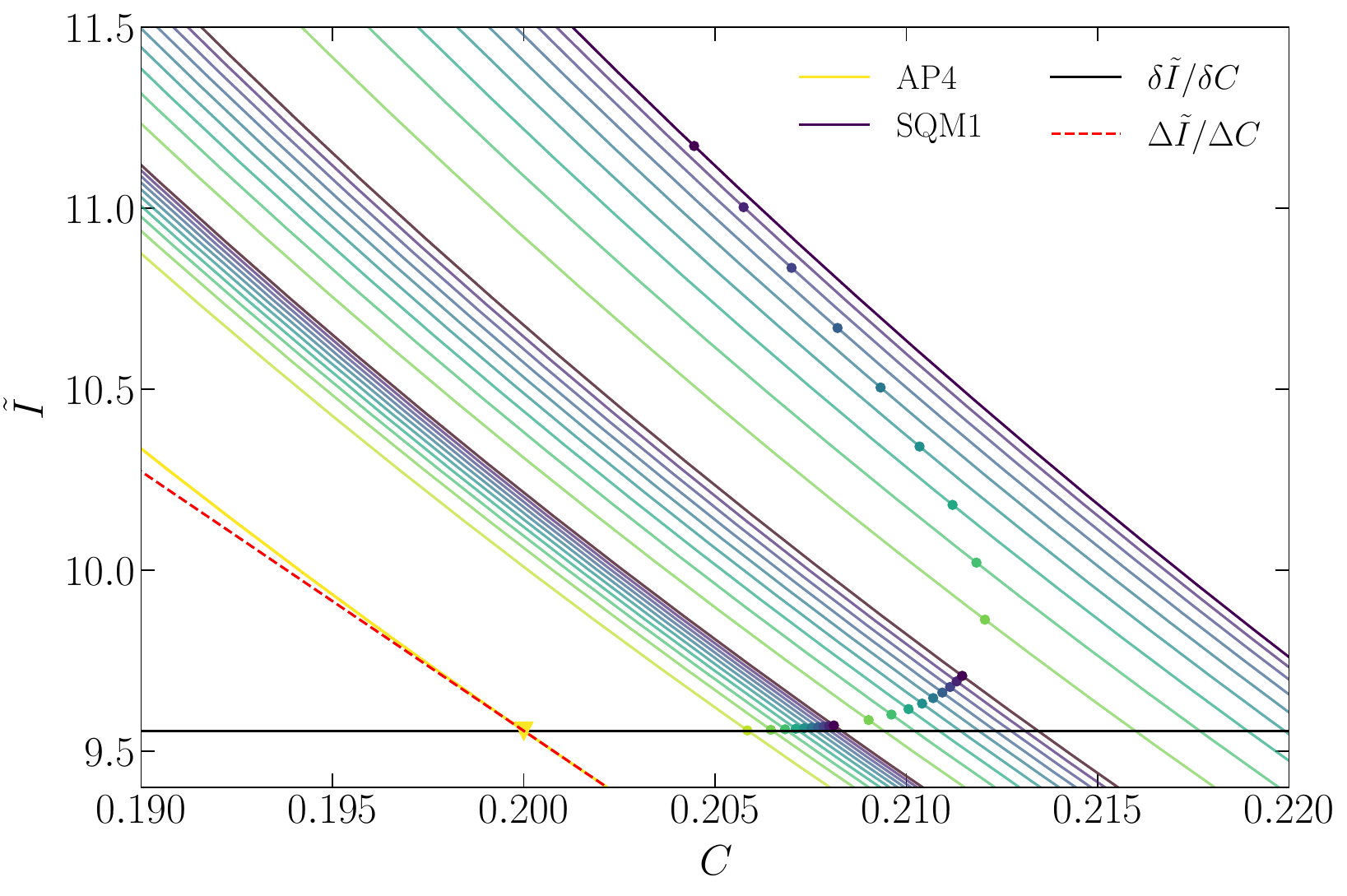}
  \caption{\label{fig:I_C_11_SQM1} Similar to Fig.~\ref{fig:I_C_11}, but for the
  SQM1 EOS. To show the deviation behavior clearly, we plot three groups of
  lines as introduced in the main text.}
\end{figure}

\begin{figure}[t]
  \centering
  \includegraphics[width=0.45\textwidth]{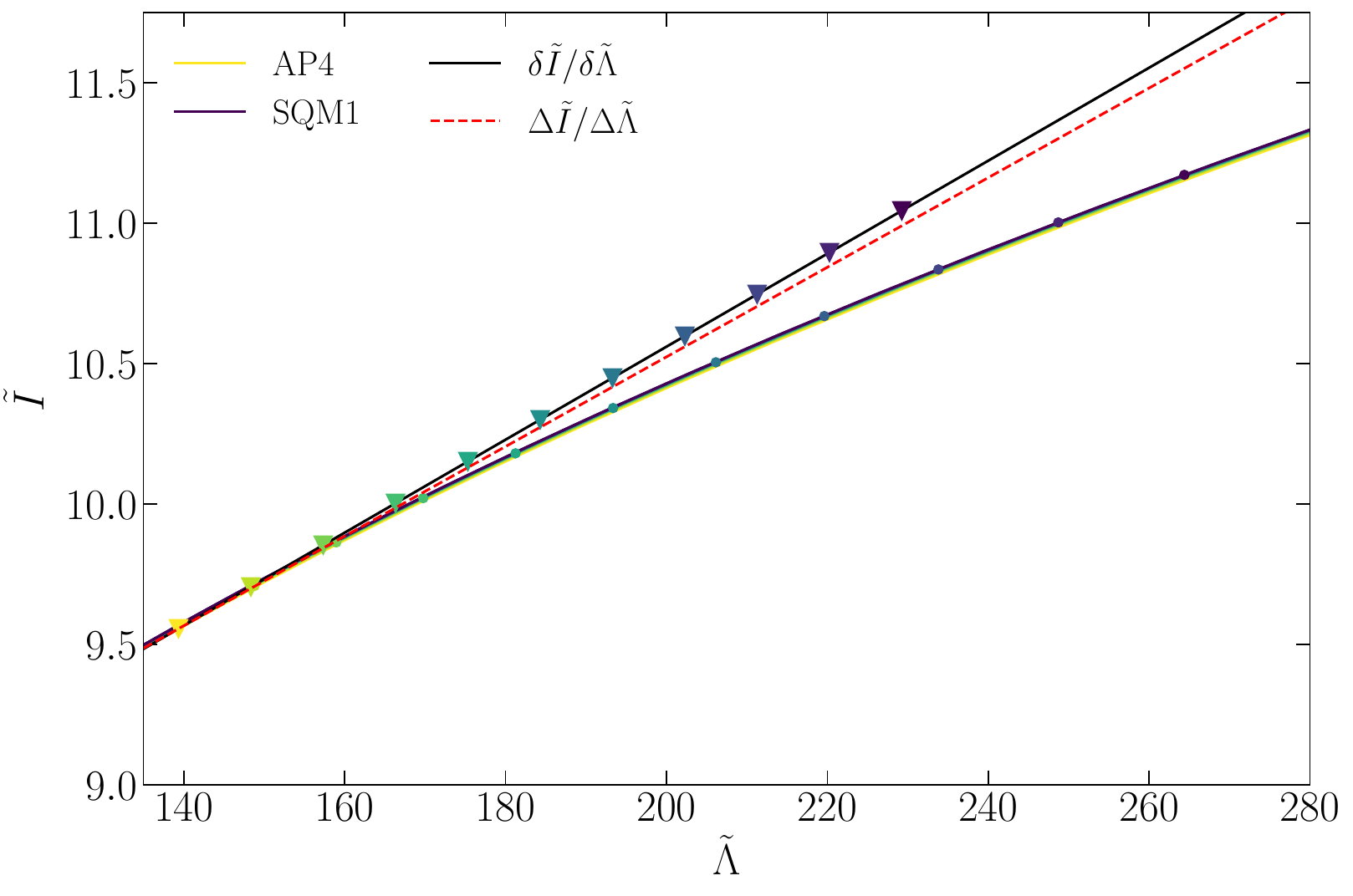}
  \caption{\label{fig:I_L_11_SQM1} Similar to Fig.~\ref{fig:I_L_11}, but for the
  SQM1 EOS.}
\end{figure}

We also show an example for NS-QS in Fig.~\ref{fig:I_C_11_SQM1} and
Fig.~\ref{fig:I_L_11_SQM1}, in which we take the EOS perturbation as the 
difference between the SQM1 EOS and the AP4 EOS. For the I-C relation, as
discussed before, from our linear theory the deviation caused by such an EOS
perturbation will diverge due to the small $p$ behavior in the
factor~(\ref{eq:DII_C}). This divergence is caused by the factor $\delta
R(p_1)/R$ and in fact leads to a divergent estimation of $\delta C$. It is also
clearly shown in Fig.~\ref{fig:I_C_11_SQM1}, which is similar to 
Fig.~\ref{fig:I_C_11} but where we plot three groups of lines for clarity. They 
correspond to EOSs that connect the AP4 EOS and the SQM1 EOS via 
\begin{eqnarray}
  \rho_{i,k}(p)&=&\rho_{\rm AP4}(p)+i\delta\rho(p)/10^k\,,\nonumber\\
    && i=0,\cdots,10\,,\quad k=1,2,3\,,
\end{eqnarray}
where $\delta\rho=\rho_{\rm SQM1}-\rho_{\rm AP4}$. We can see that, when the EOS
deviates from AP4 to SQM1, the change in the I-C relation does not depend on the
EOS perturbation linearly, which should behave like in
Fig.~\ref{fig:I_C_11_SQM1}.  The linear theory predicts this nonlinear nature,
as shown by the black line corresponding to a divergent $\delta C$, but cannot
predict the correct finite deviation that is dominated by nonlinear effects in
this case.  Starting from the QS EOS might be a more natural choice but also
cannot connect to the NS EOS.  As we discussed before, the QS EOS has a finite
density at the surface of the star so that the integral will not diverge. The
figure also shows that close to the QS, the deviation caused by the EOS 
perturbation behaves relatively normally.

The above figures all show that our linear analysis seems unable to capture the
difference between realistic EOSs as their difference is too large. However, one
aspect we have not used before is that, when considering the universal relations
among dimensionless quantities, we have a scaling freedom that corresponds to
changing the remaining unit of length in the geometric units. Namely, an EOS
with 
\begin{equation}
  \rho'(p)=\alpha\rho(p/\alpha)\,,
\end{equation}
will provide exactly the same relation as the EOS $\rho(p)$, where $\alpha$ is
any positive constant. We have already encountered this fact for polytropic EOSs
as we choose $\rho_n=1$. 
Therefore, a proper scaling may be 
applied before the linear analysis as the scaling changes the meaning of the same 
central pressure. In fact, for I-Love relation, a comparison should be made 
between the dimensionless moment of inertia of two stars with different EOSs but the 
same dimensionless tidal deformability. The quantities ${\rm D}\tilde{I}$ we 
defined in Eq.~(\ref{eq:DII_L}), originating from the rule of partial differentiation
\begin{equation}
  \left.\frac{\partial \tilde{I}(\tilde{\Lambda}(p_c,\mathrm{EOS}),\mathrm{EOS})}{\partial \mathrm{EOS}}
    \right|_{\tilde{\Lambda}}=\left.\frac{\partial \tilde{I}}{\partial{\mathrm{EOS}}}\right|_{p_c}-\left.
    \frac{\partial\tilde{I}}{\partial\tilde{\Lambda}}\right|_{\mathrm{EOS}}\left.\frac{\partial\tilde{\Lambda}}
    {\partial\mathrm{EOS}}\right|_{p_c}\,,
\end{equation} 
only provide such a comparison at the linear order. It is possible that realistic EOSs 
that largely deviate from each other might be similar after scaling. 
 Therefore, for comparing realistic EOSs, 
we suggest that a proper scaling needs to be done
before applying the linear analysis. From previous figures we find that, the
linear prediction of $\delta \tilde{I}$ is in fact pretty accurate even for
realistic EOS differences. Thus, a scaling that leads to the comparison of two
stars with a similar $C$ or $\tilde{\Lambda}$ should be well described by the
linear response.

\section{Conclusion}\label{sec:sum}

In the present paper we studied the linear response of I-C and I-Love universal
relations under arbitrary EOS perturbations. We separated the deviation of the
universal relations caused by an EOS perturbation into a product of $\delta
\rho/\rho_0$ and $p_0\rho_0{\rm D} \tilde{I}/\tilde{I}$, where the second factor
only depends on the background star configurations so that it provides an
overall estimation of the universality. We analytically calculated this factor
for a constant-density star in the Newtonian gravity, and our result for the
I-Love relation is consistent with the result in Ref.~\cite{Chan:2015iou}.  We
further applied our procedure to realistic EOSs in GR and the linear response
analysis estimates the order of magnitude of universality correctly. We found
that, for the I-C relation, the divergent behavior when comparing the QS EOS and
the NS EOS originates from the radius response $\delta R/R$, and this fully
nonlinear behavior suggests that NSs and QSs follow different I-C relations,
consistent with real scenarios. We also discussed the validity of applying the
linear analysis to study realistic EOS differences and we suggested that a
proper scaling of EOSs should be made before applying the procedure.

Our study can be regarded as a new framework for quantitative representation of
the universality for NSs. The analysis in this work may not be able to point out
the possible origin of the universality but gives an quantitative and intuitive
way to describe how good the universality is. Analytical results we obtained in
the Newtonian gravity thus can serve as a rough estimation while our analysis
can be applied in full GR cases. In this work we only focused on the I-C and
I-Love universal relations. The extension to, for example, the I-Love-Q 
relation, can be done without conceptual difficulties. Including any new
quantity basically adds one new perturbation equation that will depend on, but
not influence, the other equations as we do for $\Lambda$. We hope that our work
reported here could trigger other investigations along this direction.

\begin{acknowledgments}

We thank Kent Yagi and Takuya Katagiri for helpful discussions, and the
anonymous referee for useful comments. This work was supported by the National SKA
Program of China (2020SKA0120300), the National Natural Science Foundation of
China (124B2056, 12573042), the Beijing Natural Science Foundation (1242018), the Max
Planck Partner Group Program funded by the Max Planck Society, and the
High-Performance Computing Platform of Peking University.  Z.H.\ is supported by
the China Scholarship Council (CSC).

\end{acknowledgments}

\appendix
\section{Perturbative calculation of $\delta \Lambda/\Lambda$}
\label{app:I-L}

In this appendix, we give a relatively detailed calculation procedure for the
linear response of $\Lambda$ caused by EOS perturbations.  Starting from the
differential equation for $y$,
\begin{equation}\label{eq:app:dydp}
  \frac{{\rm d}y}{{\rm d}p}=f\left(r,m,y,\rho,\frac{{\rm d}\rho}{{\rm
  d}p}\right)\,,
\end{equation}
where we have changed the differential variable from $r$ to $p$ and have used
$f$ to denote the complex right hand side of the equation. We explicitly write
out the dependence on ${\rm d}\rho/{\rm d}p$ in $f$ as well.  Consider any EOS
perturbation $\rho'(p)=\rho(p)+\delta \rho(p)$, we can expand 
Eq.~(\ref{eq:app:dydp}) for a given background solution. Denoting 
\begin{equation}
  y'(p)=y(p)+\delta y(p)\,,
\end{equation}
at the leading order, the equation for $\delta y$ reads
\begin{eqnarray}\label{eq:app:ddydp_ori}
  \frac{{\rm d}\delta y}{{\rm d}p}&=&\frac{\partial f}{\partial r}\delta r+
  \frac{\partial f}{\partial m}\delta m +\frac{\partial f}{\partial y}\delta
  y\nonumber\\
  &&+\frac{\partial f}{\partial \rho}\delta \rho +\frac{\partial f}{\partial
  ({\rm d}\rho/{\rm d}p)}\frac{{\rm d}\delta \rho}{{\rm d}p}\,.
\end{eqnarray}
This is a linear equation in $\delta$-quantities, however,  considering $\delta
\rho=\delta(p-p_1)$, we further rewrite the equation as 
\begin{eqnarray}\label{eq:app:ddydp}
  \frac{{\rm d}\delta\tilde{y}}{{\rm d}p}&=&\frac{\partial f}{\partial r}\delta
  r+ \frac{\partial f}{\partial m}\delta m +\frac{\partial f}{\partial y}\delta
  \tilde{y} \nonumber\\
  &&+\left(\frac{\partial f}{\partial \rho}-\frac{{\rm d}}{{\rm d}p}
  \frac{\partial f}{\partial ({\rm d}\rho/{\rm d}p)}+\frac{\partial f}{\partial
  y} \frac{\partial f}{\partial ({\rm d}\rho/{\rm d}p)}\right)\delta \rho\,,
\end{eqnarray}
where 
\begin{equation}
  \delta\tilde{y}\equiv\delta y-\frac{\partial f}{\partial ({\rm d}\rho/{\rm
  d}p)}\delta \rho\,,
\end{equation}
and we have initial condition $\delta y \big|_{p=p_0} =0$.
Equation~(\ref{eq:app:ddydp}) now can be solved with a similar procedure we
introduced in Sec.~\ref{sec:local U}. Considering $\delta\rho=\delta(p-p_1)$, we
can solve for $\delta \tilde{y}$ with 
\begin{equation}
  \frac{{\rm d}\delta\tilde{y}}{{\rm d}p}=\frac{\partial f}{\partial r}\delta r+
  \frac{\partial f}{\partial m}\delta m +\frac{\partial f}{\partial y}\delta
  \tilde{y}\,,
\end{equation}
and initial conditions
\begin{equation}
  p=p_1\,,\quad \delta\tilde{y}=-\left.\left(\frac{\partial f}{\partial
  \rho}-\frac{{\rm d}} {{\rm d}p} \frac{\partial f}{\partial ({\rm d}\rho/{\rm
  d}p)}+\frac{\partial f}{\partial y} \frac{\partial f}{\partial ({\rm
  d}\rho/{\rm d}p)}\right)\right|_{p=p_1}\,.
\end{equation}
We shall note that, one can directly integral Eq.~(\ref{eq:app:ddydp_ori}) 
around $p_1$ to obatin the integration initial condition for $\delta y$ in 
$p=p_1^-$ that is equivalent to what we have done here. However, one should 
be careful that $\delta y$ contains a delta function part that also 
contributes to the jump condition, which is the last term in the above 
equation.

For calculating $\delta \Lambda$, we are interested in the value at the
boundary, $\delta Y=\delta y \big|_{p=0^-}$, where we use $0^-$ to denote the
point outside the star but infinitely close to the surface. One needs to be
careful with the fact that, similarly to the calculation of the QS tidal Love
number, the ${\rm d}\rho/{\rm d}p$ term in the equation leads to additional
correction terms at the boundary. For example, the relevant term that is
proportional to $\delta r$ reads
\begin{equation}
  \frac{{\rm d}\delta {y}}{{\rm d}p}=\cdots+\frac{12\pi r^3 m}{(m+4\pi r^3p)^2}
  \frac{\delta r}{r}\frac{{\rm d}\rho}{{\rm d}p}+\cdots\,,
\end{equation}
which leads to a correction term 
\begin{equation}
  \delta {y} \big|_{p=0^-} - \delta {y}\big|_{p=0^+}=-\frac{12\pi
  R^3\rho_s}{M}\frac{\delta R}{R}+\cdots\,,
\end{equation}
where $\rho_s$ is the density at the surface of the star and dots represent the
corrections come from other terms. We have used $0^+$ to denote the point that
is inside but infinitely close to the surface opposite to $0^-$.



With $\delta Y$ in hand, $\delta \Lambda$ then can be calculated via 
\begin{equation}
  \delta \Lambda=\frac{\partial \Lambda}{\partial M}\delta M+
  \frac{\partial \Lambda}{\partial R}\delta R+
  \frac{\partial \Lambda}{\partial Y}\delta Y\,,
\end{equation}
where $\Lambda$ given by Eq.~(\ref{eq:k2L}) and Eq.~(\ref{eq:Y2k}) is regarded
as a function of $M$, $R$, and $Y$.

\bibliography{refs.bib}

\end{document}